\documentclass{article}
\usepackage{graphicx}

\begin{document}

\title{ Pad\'e-Summation Approach to QCD $\beta$-Function Infrared Properties}

\author{F. A. Chishtie,$^1$
 V. Elias,$^1$  V.~A.~Miransky,$^{1,2,3}$ 
~T.~G.~ Steele$^4$
}
\footnotetext[1]{Department of Applied Mathematics,
The University of Western Ontario,
London, Ontario N6A 5B9 CANADA}
\footnotetext[2]{Department of Physics,
Nagoya University,
Nagoya 464-8602 JAPAN}
\footnotetext[3]{Permanent Address:  
Bogolyubov Institute for Theoretical Physics, Ukranian National Academy of Sciences, 
UA 252143 Kiev, UKRAINE}
\footnotetext[4]{Department of Physics \& Engineering Physics,
University of Saskatchewan,
Saskatoon, Saskatchewan S7N 5C6 CANADA}

\maketitle

\begin{abstract}
     In view of the successful asymptotic Pad\'e-approximant predictions for higher-loop terms
within QCD and massive scalar field theory, we address whether Pad\'e-summations of the 
$\overline{MS}$ QCD $\beta$-function for a given number of flavours exhibit an infrared-stable fixed point, or
alternatively, an infrared attractor of a double valued couplant as noted by Kogan and Shifman for
the case of supersymmetric gluodynamics. Below an approximant-dependent flavour threshold 
$(6 \leq n_f \leq 8)$, we find that Pad\'e-summation $\beta$-functions incorporating $[2|1], [1|2], [2|2], [1|3]$, and $[3|1]$
approximants whose Maclaurin expansions match known higher-than-one-loop contributions to the
$\beta$-function series always exhibit a positive pole prior to the occurrence of their first positive zero,
precluding any identification of this first positive zero as an infrared-stable fixed point of the $\beta$-
function. This result is shown to be true regardless of the magnitude of the presently-unknown 
five-loop
$\beta$-function contribution explicitly appearing within Pad\'e-summation $\beta$-functions incorporating
$[2|2], [1|3]$, and $[3|1]$ approximants. Moreover, the pole in question suggests the occurrence of
dynamics in which both a strong and an asymptotically-free phase share a common
infrared attractor. We briefly discuss the possible relevance of
infrared-attractor dynamics to the success of recent calculations of the
glueball mass spectra in QCD with $N_c \rightarrow \infty$ via
supergravity.
As $n_f$ increases above an approximant-dependent flavour threshold, 
Pad\'e-summation $\beta$-functions incorporating $[2|2], [1|3]$, and $[3|1]$ approximants exhibit dynamics
controlled by an infrared-stable fixed point over a widening domain of the five-loop $\overline{MS}$  $\beta$-function
parameter $(\beta_4/\beta_0)$. Subsequent to the above-mentioned flavour threshold, all approximants considered
exhibit infrared-stable fixed points that decrease in
magnitude with increasing flavour number.
\end{abstract}

\section{Introduction}

Asymptotic Pad\'e-approximant methods have been utilized to estimate
higher order contributions to renormalization-group (RG) functions
within both QCD \cite{EJJ,EKS,ESC} and massive scalar field theory
\cite{EKS,ESC,CES}, for which such estimates compare quite favourably
with explicit calculation \cite{KNS}.  More recently, such methods have
been shown to predict RG-accessible coefficients of logarithms within
five-loop-order contributions to QCD correlation functions with striking
accuracy \cite{CE}.  These results are all derived from an improvement of
Pad\'e-estimated coefficients which incorporates the estimated error of
Pad\'e-approximants in predicting the $n!K^{-n}n^\gamma$ asymptotic
behaviour expected for $n^{th}$ order coefficients of a field-theoretic
perturbative series \cite{VZ,SEK}.  For $[N|M]$ approximants, such error
is seen to decrease with increasing N and M \cite{EJJ,EKS,VZ}, as well
as to favour diagonal and near-diagonal approximants \cite{EG}.

Of course, the use of such higher approximants becomes tenable only if
the corresponding perturbative series is known to sufficiently high
order.  A known series of the form $\sum_{j=1}^k R_j x^j$ specifies all
coefficients within $[N|M]$-approximants to the series only for
$\{N,M\}$ such that $k=N+M$;  even four-loop calculations (corresponding to $k=3$ if
the leading $x^2$ behaviour is factored out from the series) serve only to
specify $[2|1]$, $[1|2]$, and $[0|3]$ approximants.  Nevertheless, such
approximants have been used in conjunction with the anticipated
asymptotic error to predict the next (five-loop) coefficient $R_4$ as
well as the corresponding diagonal $[2|2]$ approximant to the full
field-theoretic series.\cite{EJJ,EKS,ESC,CES,CE}

The success of such predictions for those cases in which $R_4$ is known
\cite{EJJ,ESC,CE} suggests that

\begin{enumerate}
\item Pad\'e approximants determined from $\{R_1 , ... , R_k\}$ more accurately
represent the field-theoretic asymptotic series $\sum_{j=0} R_j x^j$ than mere 
truncation of this series to $\sum_{j=0}^k R_j x^j$;   
\item Sufficiently-high Pad\'e-approximants grow arbitrarily close to
the function of $x$ represented by the full field-theoretic series, as
suggested by asymptotic error formulae \cite{EJJ} following from
renormalon-estimates of large-$j$ coefficients in the series \cite{VZ}.
\end{enumerate}
In the absence of alternatives other than explicit series truncation,
such ``Pad\'e-summation'' \cite{EJJ,EKS} of the full perturbative series
may provide a much wanted means for extrapolating such series to the
infrared region.  We are particularly interested in two possible scenarios for infrared
dynamics within QCD, either the infrared attractor suggested by Kogan and Shifman
within the context of supersymmetric gluodynamics \cite{KS}, or alternatively, dynamics
governed by an infrared-stable fixed point. In reference \cite{EKS}, for example, a 
Pad\'e-summation of the $n_f = 3$ QCD $\beta$-function is argued to contain a
zero corresponding to an infrared fixed point comparable to that
predicted by Mattingly and Stevenson \cite{MS,ACM}.  

Indeed it is this claim that provides some of the motivation for our present
work.  In the approach of ref. \cite{ACM}, an infrared-stable fixed
point is argued to occur even when $n_f = 0$, in contradiction to a
broadening consensus that values of $n_f$ even larger than 3 are
required for infrared-stable fixed points to occur \cite{ATW,MY,RSC,VEL,EGM}; 
e.g. $n_f = 8$
is suggested by a two-loop truncation of the QCD $\beta$-function, with
consequences for the phase-structure of QCD first explored by Banks and
Zaks \cite{BZ}.

In the present paper, we utilize the perturbative series for the
$\overline{MS}$ QCD $\beta$-function, now known in full to four-loop
order \cite{RVL}, in order to construct $[N|M]$ Pad\'e summations, which
are assumed to provide information about the $\beta$-function's first 
positive zero or pole (this
point is further discussed in Section 2).  We are able to extend our
analysis past $N+M=3$ by expressing $N+M=4$ approximants in terms of the
presently-unknown five-loop $\beta$ function coefficient, which is
treated here as a variable parameter.  Among the specific issues we
address in the sections that follow are:

\begin{enumerate}
\item the existence of a flavour-threshold for dynamics governed by an infrared 
stable-fixed point,
\item whether differing Pad\'e-approximants are consistent in predicting
infrared properties of QCD,
\item the dependence of Pad\'e-predictions for $\beta$-function infrared properties 
on the presently-unknown five-loop term,
\item the size of the infrared fixed point, particularly in
comparison with the $\alpha_s^* = \pi/4$ benchmark value for chiral-symmetry 
breaking \cite{ATW,MY},
\item the existence of a strong phase of QCD for $n_f = 0$ \cite{KS}
and, possibly, for nonzero $n_f$ as well, and
\item the elevation of the true infrared cutoff (mass gap) of $n_f = 3$ QCD to hadronic
mass scales (500 - 700 MeV).
\end{enumerate}

In Section 2, we discuss how Pad\'e-approximants constructed from the
known terms of the $\beta$-function series can exhibit information about
the infrared behaviour of the corresponding couplant.  This approach
relies upon the Pad\'e-approximant remaining closer to the true 
$\beta$-function than the truncated perturbation series from which the
approximant is constructed, as discussed above.  We conclude Section 2
by obtaining Pad\'e-summation expressions for QCD $\overline{MS}$
$\beta$-functions which incorporate $[2|1]$, $[1|2]$, $[2|2]$, $[1|3]$ and
$[3|1]$ approximants to post-one-loop terms in the $\beta$-function
series.  The latter three approximants are expressed in terms of a
variable $R_4 (\equiv \beta_4 / \beta_0)$ characterizing the presently
unknown 5-loop contribution to the $\overline{MS}$ $\beta$-function.

In Section 3, we apply such Pad\'e-summation methods to the 
$\beta$-function characterizing $N_c = 3 \; QCD$ with no fundamental fermions.
Pad\'e-summation  predictions for the infrared structure of $n_f = 0$ QCD in the 
't Hooft ($N_c \rightarrow \infty$) limit \cite{GTF} are presented separately in an Appendix. 
For both cases, we find that no Pad\'e-summation $\beta$-function supports the existence of an 
infrared-stable fixed point for the $n_f = 0$ QCD couplant.  Moreover, we
demonstrate that the infrared behaviour extracted from Pad\'e-
summations of the $n_f = 0$ QCD $\beta$-function appears to be governed
by an apparent $\beta$-function pole, an infrared-attractor of { \it two}
ultraviolet phases of the couplant.  This behaviour is in qualitative
agreement with that extracted from supersymmetric QCD in the absence of
fundamental-representation matter fields \cite{KS}. We conclude
Section 3 with a brief discussion of the possible
applicability of infrared-attractor dynamics to the glueball
spectrum for the $N_c \rightarrow \infty$ case. 

In Section 4, we extend the analysis of Section 3 to nonzero $n_f$.
Specifically, we examine Pad\'e-summation $\beta$-functions which
incorporate $[2|1]$, $[1|2]$, $[2|2]$, $[1|3]$ and $[3|1]$ approximants
to post-one-loop terms in the perturbative $\beta$-function series.  We
find that all such approximants exhibit a flavour threshold for the
occurrence of infrared dynamics characterized by an infrared-stable
fixed point.  Beneath this threshold, which occurs between 6 and 9
flavours (depending on the approximant), no infrared-stable fixed point
is possible {\it regardless of the magnitude of the unknown five-loop
term}
($\beta_4$) entering such approximants.  Above the threshold, we observe
a progressively broadening domain of $\beta_4/\beta_0$ for which an
infrared-stable fixed point occurs, as well as a decrease in the
magnitude of such fixed points with increasing $n_f$.

In Section 5, we focus on the infrared behaviour of the $n_f
= 3$ case.  We show that $[2|2]$, $[1|3]$ and $[3|1]$ Pad\'e-summations
of the $n_f = 3$ perturbative $\beta$-function series yield similar
infrared dynamics to the ``gluodynamic'' $n_f = 0$ case of Section 3.
Such summations are all shown to yield an enhanced mass gap, an infrared boundary to the
domain of $\alpha_s$ somewhat in excess of 500 MeV, regardless of the 5-loop 
contribution to the $\beta$-function series.  This infrared boundary is shown
to be remarkably stable against such 5-loop
corrections to the $\beta$-function.

\section{Methodology}
\subsection{\it A Toy $\beta$-Function:}

\renewcommand{\theequation}{2.\arabic{equation}}
\setcounter{equation}{0}

Pad\'e-approximants to a function whose Maclaurin series is $1 +
\sum_{k=1} R_k x^k$ are well known to be valid for a broader range of
the expansion parameter $x$ than truncations of the series.
Consider, for example, the following toy $\beta$-function
\begin{equation}
\mu^2 \frac{dx}{d\mu^2} = \beta_A (x) \equiv -x^2 [sec(x) - tan(x)]
\end{equation}
We have chosen $\beta_A$ to be asymptotically free, i.e., to have an
ultraviolet fixed point at $x = 0$.  Since
\begin{equation}
\lim_{x \rightarrow \pi/2} \left( sec (x) - tan (x) \right) = 0
\end{equation}
we have also chosen $\beta_A$ to have an infrared fixed point at $x =
\pi/2$.  $\beta_A$ has a subsequent pole at $x = 3 \pi/2$, and $\beta_A$
alternates zeros and poles as $x$ increases by subsequent increments of
$\pi$.  The point here, however, is that the solution to (2.1) will
exhibit the same dynamics as depicted in Fig. 1 for $x$ between zero and $\pi/2$,
corresponding to a freezing-out of the coupling at the $x = \pi/2$
infrared-stable fixed point.

Suppose, however, that the sum-total of our knowledge of $\beta_A$ is
the first five-terms of this series expansion, corresponding to a
hypothetical five-loop $\beta$-function calculation:
\begin{equation}
\beta_A^{(5)} (x) = -x^2 \left[ 1 - x + \frac{x^2}{2} - \frac{x^3}{3} +
\frac{5x^4}{24} \right]
\end{equation}
This truncated series is, of course, not equal to zero at the $x =
\pi/2$ infrared fixed point.  Rather, when $x = \pi/2$, each term in the
series is seen to be comparable to prior lower-order terms:
\begin{equation}
\beta_A^{(5)}(\pi/2) = - \frac{\pi^2}{4} \left[ 1 -
1.571 + 1.233 - 1.292 + 1.269 \right]
\end{equation}
One would necessarily conclude that the field
theoretical calculation leading to (2.3) cannot be extended to large
enough $x$ to extract information about the infrared properties of
$\beta_A (x)$.

The series (2.3), however, provides sufficient information to construct
a $[2|2]$ approximant to the degree-four polynomial within
(2.3):

\begin{equation}
\beta_A^{[2|2]} (x) = -x^2 \left[ \frac{1 - \frac{x}{2} -
\frac{x^2}{12}}{1 + \frac{x}{2} - \frac{x^2}{12}}\right]
\end{equation}
Equation (2.5) is obtained by requiring that the degree-2 numerator and
denominator polynomials of the $[2|2]$ approximant be chosen so as to
yield a Maclaurin expansion whose first five terms reproduce the five terms in (2.3).
One can easily verify that $\beta_A^{[2|2]}$ remains closer to $\beta_A$
(2.1) over a much larger range of $x$ than $\beta_A^{(5)}$, as given in (2.3).  This
range is inclusive of the first zero of $\beta_A$.  $\beta_A^{[2|2]}$
has a positive zero at $x = 1.583$, quite close to $\beta_A$'s true
zero at $\pi/2$.  Moreover, the denominator in (2.5) remains positive
over the entire range $0 \leq x \leq 1.583$, guaranteeing that $x =
1.583$ is infrared-stable (a sign change would render this fixed-point
ultraviolet-stable).  Thus, Pad\'e-improvement of the information in
(2.3) provides a means for extracting information about the infrared
properties of $\beta_A$ that is otherwise inaccessible from the
``five-loop'' expression.

\subsection{\it $\beta$-Function Poles}

It should also be noted that (2.5) predicts that a pole at $x = 7.58$
follows the zero at 1.583 without a second intervening zero.  This
result is qualitatively similar to the true behaviour of $\beta_A$,
which acquires a pole at $x = 3\pi/2$ subsequent to the zero at $\pi/2$
without any additional intervening
zeros.  However, accuracy in predicting
this pole, as well as any subsequent zeros or poles, is clearly beyond
the scope of (2.5), the $[2|2]$ Pad\'e-summation of $\beta_A$.  

We have seen, however, that Pad\'e methods do provide a window for viewing 
leading $\beta$-function singularities that would otherwise be inaccessible.
One cannot automatically dismiss the possibility of such singularities occurring
within QCD $\beta$-functions.  For example, the $\beta$-function of
$SU(N_{c})$ SUSY
gluodynamics, which is
known {\it exactly} if no matter fields are present, exhibits precisely
such a zero \cite{NSV}:
\begin{equation}
\beta(x) = -\frac{3N_{c}x^2}{4} \left[ \frac{1}{1-N_{c}x/2}
\right]; \; \;x
\equiv \frac{\alpha_s}{\pi}.
\end{equation}
Eq. (2.6), which can be derived via imposition of the Adler-Bardeen theorem
upon the supermultiplet of the anomalies \cite{DRT}, implies the existence of a
strong ultraviolet phase (the upper branch of Fig. 2) when the couplant
$x$ is greater than the $\beta$-function pole at $2/N_{c}$ \cite{KS}.
Interestingly, the $\beta$-function (2.6) is itself a $[0|1]$
approximant once the leading $-3N_{c}x^2/4$ coefficient is factored out.

To demonstrate how Pad\'e summation provides a window for extracting possible
pole singularities in true $\beta$-functions, we consider a second toy
example
\begin{equation}
\mu^2 \frac{dx}{d\mu^2} \equiv \beta_B (x) = x^2 \left[ sec (x) + tan
(x) \right].
\end{equation}
$\beta_B (x)$ is asymptotically free, but has a positive {\it pole} at $x =
\pi/2$ prior to its first zero at $x = 3\pi/2$.  This zero is an {\it
ultraviolet} stable fixed point because of the overall sign change
associated with passing through the pole at $x = \pi/2$.  The behaviour
of $x(\mu)$ for $x < 3\pi/2$ is schematically depicted in Fig 2,
with $x = \pi/2$ corresponding to $\mu_c$, the minimum allowed value
of $\mu$ (assuming the couplant $x$ is real).  

Such infrared structure is not at all evident in
the ``five-loop'' approximation to (2.7)
\begin{equation}
\beta_B^{(5)} (x) = -x^2 \left[ 1 + x + \frac{x^2}{2} + \frac{x^3}{3} +
\frac{5x^4}{24} \right],
\end{equation}
an expression which ceases to be close to the true $\beta$-function
(2.7) for values of $x$ substantially smaller than $x = \pi/2$.
However, one can obtain a $[2|2]$ approximant directly from the
truncated series in (2.8)
\begin{equation}
\beta_B^{[2|2]} (x) = -x^2 \frac{[1+\frac{x}{2} - \frac{x^2}{12}]}{[1 -
\frac{x}{2} - \frac{x^2}{12}]}
\end{equation}
whose Maclaurin expansion yields (2.8) for its first five terms.
The first denominator zero of (2.9) is at $x = 1.583$, in good agreement
with the positive pole of (2.7) at $x = \pi/2$.  Moreover, the first
denominator zero of (2.9) precedes all (positive) numerator zeros,
thereby eliminating the possibility of the pole being preceded by an
infrared fixed point.  Thus, (2.9) and the true $\beta$-function (2.7)
predict very similar dynamics between the ultraviolet-stable fixed point
at $x = 0$ and the infrared-attractor {\it pole} at $x = \pi/2$.  By contrast,
the ``five-loop'' $\beta$-function (2.8) can only reproduce the true
running couplant in the ultraviolet region where $x$ is near zero.

\subsection{\it Pad\'e-Improvement and Infrared Behaviour}

It is to be emphasized that the examples presented above demonstrate how
Pad\'e-improvement {\it may} provide information about the infrared region
that a truncated perturbative series cannot.  There is no way, of
course, to prove that the first positive zero or pole of a given 
Pad\'e-summation
is indeed the first zero or pole of the true $\beta$-function.
In the absence of methodological alternatives, however, we will explore
below the consequences of {\it assuming} this to be the case.
Corroboration of such an assumption relies ultimately on an explicit
comparison of next-order terms calculated for the $\overline{MS}$ QCD
$\beta$-function series, to Pad\'e-predictions for these terms ({\it e.g.},
ref \cite{ESC}).  There is reason to be encouraged, however, by the
success already demonstrated for Pad\'e predictions of the known five-loop
content of the massive scalar field theory $\beta$-function
\cite{EJJ,CES}.  Similar successes are obtained in predicting 
RG-accessible coefficients within the five loop contributions to QCD vector
and scalar fermionic-current correlation functions, as well as within
four-loop contributions to the scalar gluonic-current correlation
function \cite{ESC,CE}.

The general approach we take is to replace a $k$-loop truncation of the
asymptotic $\beta$-function series
\begin{equation}
\beta^{(k)} (x) = -\beta_0 x^2 \left( 1 + R_1 x + ... R_{k-1} x^{k-1}
\right)
\end{equation}
with an expression incorporating the corresponding $[N|M]$ Pad\'e-
approximant $(N+M = k-1)$:
\begin{equation}
\beta^{[N|M]} (x) = -\beta_0 x^2 \left( \frac{1+a_1 x + ... + a_N
x^N}{1+b_1 x + ... + b_M x^M} \right)
\end{equation}
The $N+M$ coefficients $\{a_1, ... ,a_N, b_1, ... ,b_M\}$ are completely
determined by the requirement that the first $k$ terms in the Maclaurin
expansion of (2.11) replicate $\beta^{(k)} (x)$ (2.10).  We then examine
$\beta^{[N|M]}$ in order to determine whether or not it is supportive of
an infrared-stable fixed point, as in Fig. 1, or an infrared-attractor
pole, as in Fig. 2.

If the first positive zero of the degree-N polynomial in the numerator
of (2.11) precedes any positive zeros of the degree-M polynomial in the
denominator, that first numerator zero corresponds to the 
infrared-stable fixed point at which the couplant $x$ freezes out in Fig. 1.
Alternatively, if the first positive zero in the denominator of (2.11)
precedes any positive zeros in the degree-M numerator polynomial, that
first denominator zero corresponds to the infrared-attractor pole common
to both couplant phases of Fig. 2.

\subsection{\it A Pad\'e Roadmap}

It will prove useful to tabulate those formulae required to obtain
Pad\'e approximants (2.11) whose Maclaurin expansions reproduce the
truncated series (2.10) for the $\overline{MS}$ $\beta$-function.
Values for $\beta_0$, $R_1$, $R_2$ and $R_3$ for the $\beta$-function,
as defined by (2.10), are tabulated in Table \ref{tab1}.  Corresponding $[2|1]$
and $[1|2]$ approximants to the truncated series $1 + R_1 x + R_2 x^2 +
R_3 x^3$ within (2.10) are given by the following formulae:

\begin{table}[htb]
\begin{center}
\begin{tabular}{ccccc}
$n_f$ & $\beta_0$ & $R_1$ & $R_2$ & $R_3$\\
\hline \\
0 & 11/4 & 51/22 & 2857/352 & 41.5383\\
1 & 31/12 & 67/31 & 62365/8928 & 34.3295\\
2 & 29/12 & 115/58 & 48241/8352 & 27.4505\\
3 & 9/4 & 16/9 & 3863/864 & 20.9902\\
4 & 25/12 & 77/50 & 21943/7200 & 15.0660\\
5 & 23/12 & 29/23 & 9769/6624 & 9.83592\\
6 & 7/4 & 13/14 & -65/224 & 5.51849\\
7 & 19/12 & 10/19 & -12629/5472 & 2.42409\\
8 & 17/12 & 1/34 & -22853/4896 & 1.00918\\
9 & 5/4 & -3/5 & -1201/160 & 1.97366\\
10 & 13/12 & -37/26 & -41351/3744 & 6.44815\\
11 & 11/12 & -28/11 & -49625/3168 & 16.3855\\
12 & 3/4 & -25/6 & -6361/288 & 35.4746\\
13 & 7/12 & -47/7 & -64223/2016 & 71.6199\\
14 & 5/12 & -113/10 & -70547/1440 & 145.373\\
15 & 1/4 & -22 & -2823/32 & 332.091\\
16 & 1/12 & -151/2 & -81245/288 & 1309.98
\end{tabular}
\caption{The known coefficients $\beta_0$ and $R_{1,2,3}$ appearing in the QCD
$\beta$-function
(2.10) for $n_f = 0 - 16$, as calculated in \cite{RVL}.}
\label{tab1}
\end{center}
\end{table}

\renewcommand{\theequation}{2.12\alph{equation}}
\setcounter{equation}{0}

\begin{equation}
\beta^{[2|1]} (x) = -\beta_0 x^2 \left[ \frac{1+a_1 x + a_2 x^2}{1+b_1
x} \right],
\end{equation}
\begin{equation}
a_1 = (R_1 R_2 - R_3) / R_2,
\end{equation}
\begin{equation}
a_2 = (R_2^2 - R_1 R_3)/R_2,
\end{equation}
\begin{equation}
b_1 = -R_3 / R_2,
\end{equation}
and
\renewcommand{\theequation}{2.13\alph{equation}}
\setcounter{equation}{0}
\begin{equation}
\beta^{[1|2]} (x) = -\beta_0 x^2 \left[ \frac{1+a_1 x}{1 + b_1 x + b_2
x^2} \right],
\end{equation}
\begin{equation}
b_1 = (R_3-R_1 R_2) / (R_1^2 - R_2),
\end{equation}
\begin{equation}
b_2 = (R_2^2 - R_1 R_3) / (R_1^2 - R_2),
\end{equation}
\begin{equation}
a_1 = R_1 + b_1.
\end{equation}
We do not consider the $[0|3]$-approximant, as this approximant 
has no possible numerator zeros and therefore cannot
lead to an infrared fixed point.  The $[3|0]$-approximant is, of
course, the truncated series itself.  The ``diagonal-straddling''
$[1|2]$ and $[2|1]$ approximants are the only $N+M = 3$ approximants for
which both infrared-stable fixed points (Fig. 1) and infrared-attractor
poles (Fig. 2) are possible, depending on the specific ordering of
positive numerator and denominator zeros in (2.12a) and (2.13a).  Hence,
it is these approximants we will study in subsequent sections.

$R_4$, the ``next-order'' coefficient of the QCD $\beta$-function (2.10),
is not presently known.  Nevertheless, one can construct 
``diagonal-straddling'' $[2|2]$, $[1|3]$ and $[3|1]$ approximants with $R_4$ 
taken to be an
arbitrary  parameter, by utilizing the values for $\{R_1, R_2, R_3\}$ given
in Table \ref{tab1}. One finds for the truncated series $\beta^{(5)}(x)$ $= -
\beta_0 x^2 \left[ 1+R_1 x + R_2 x^2 + R_3 x^3 + R_4 x^4 \right]$ the
following Pad\'e-approximant $\beta$-functions:

\renewcommand{\theequation}{2.14\alph{equation}}
\setcounter{equation}{0}
\begin{equation}
\beta^{[2|2]} (x) = -\beta_0 x^2 \left[ \frac{1 + a_1 x + a_2 x^2}{1+b_1
x + b_2 x^2} \right],
\end{equation}
\begin{equation}
b_1 = (R_1 R_4 - R_2 R_3) / (R_2^2 - R_1 R_3),
\end{equation}
\begin{equation}
b_2 = (R_3^2 - R_2 R_4) / (R_2^2 - R_1 R_3),
\end{equation}
\begin{equation}
a_1 = R_1 + b_1,
\end{equation}
\begin{equation}
a_2 = R_1 b_1 + b_2 + R_2;
\end{equation}

\renewcommand{\theequation}{2.15\alph{equation}}
\setcounter{equation}{0}
\begin{equation}
\beta^{[1|3]} (x) = -\beta_0 x^2 \left[ \frac{1+a_1 x}{1+b_1 x + b_2 x^2
+ b_3 x^3} \right]
\end{equation}
\begin{equation}
b_1 = \frac{(R_1^2 R_2 - R_2^2 - R_1 R_3 + R_4)}{(2R_1 R_2 - R_1^3 -
R_3)},
\end{equation}
\begin{equation}
b_2 = \frac{(R_1^2 R_3 - R_1 R_2^2 + R_2 R_3 - R_4 R_1)}{2R_1 R_2 -
R_1^3 - R_3},
\end{equation}
\begin{equation}
b_3 = \frac{[R_3^2 + R_2^3 - 2R_1 R_2 R_3 + R_4(R_1^2 - R_2)]}{2R_1 R_2
- R_1^3 - R_3}
\end{equation}
\begin{equation}
a_1 = R_1 + b_1;
\end{equation}

\renewcommand{\theequation}{2.16\alph{equation}}
\setcounter{equation}{0}
\begin{equation}
\beta^{[3|1]} (x) = -\beta_0 x^2 \left[ \frac{1+a_1 x + a_2 x^2 + a_3
x^3}{1+b_1 x} \right],
\end{equation}
\begin{equation}
a_1 = R_1 - R_4/R_3,
\end{equation}
\begin{equation}
a_2 = R_2 - R_1 R_4 / R_3,
\end{equation}
\begin{equation}
a_3 = R_3 - R_2 R_4 / R_3,
\end{equation}
\begin{equation}
b_1 = -R_4 / R_3.
\end{equation}

Given the known values of $\{R_1, R_2, R_3\}$ tabulated in Table \ref{tab1}, we
have tabulated $\beta^{[2|2]}$'s coefficients $\{a_1, a_2, b_1, b_2\}$
in Table \ref{tab2} for all $n_f$-values for which $x=0$ is an ultraviolet-stable fixed point.  
These coefficients are all linear in the unknown parameter $R_4$.  A
similar tabulation of the coefficients within $\beta^{[1|3]}$ and
$\beta^{[3|1]}$ is presented in Tables \ref{tab3} and \ref{tab4}, respectively.  Because
all such Pad\'e-coefficients are $R_4$-dependent for a given choice of $n_f$, 
one might expect to find infrared stable fixed point behaviour (Fig. 1) for
some range of $R_4$, infrared-attractor pole behaviour (Fig. 2) for
another range of $R_4$, and (possibly) some regimes of $R_4$ for which
there are neither numerator nor denominator zeros.  Surprisingly, we
find in Section 4  that infrared-stable fixed point behaviour as
in Fig. 1 does not occur {\it for any} of the diagonal-straddling
approximants (2.12-15) until $n_f$ reaches a threshold value, 
{\it regardless of $R_4$}.  This
threshold depends on the approximant considered, but is greater than or
equal to 6 for all $[N|M]$ approximants discussed above.

\begin{table}[htb]
\small
\begin{center}
\begin{tabular}{ccccc}
$n_f$ & $a_1$ & $a_2$ & $b_1$ & $b_2$\\
\hline\\
0& 13.4026-0.0762155 $R_4$ & -22.9153+0.0901663$R_4$ & -0.0762155$R_4$ + 11.0844 &
0.266848$R_4$ - 56.7275\\
1 & 116019-0.0850858$R_4$ & -19.0068+0.0911037$R_4$ & -0.0850858$R_4$+9.44058 &
0.274999$R_4$-46.3959\\
2 & 9.50934-0.0941220$R_4$ & -15.0709+0.0875660$R_4$ & -0.0941220$R_4$+7.52659 &
0.274187$R_4$-35.7703\\
3 & 7.19456-0.102610$R_4$ & -11.3292+0.0756438$R_4$ & -0.102610$R_4$ + 5.41678 &
0.258062$R_4$-25.4301\\
4 & 4.84008-0.110684$R_4$ & -8.18415+0.0485887$R_4$ & -0.110684$R_4$+3.30008 &
0.219042$R_4$-16.3139\\
5 & 2.67929-0.123291$R_4$ & -6.19674-0.0112453$R_4$ & -0.123291$R_4$+1.41842 &
0.144208$R_4$ - 9.45997\\
6 & 0.610851-0.184236$R_4$ & -6.62748-0.228650$R_4$ & -0.184236$R_4$-0.317721 &
-0.0575739$R_4$-6.04228\\
7 & 1.907466+0.129932$R_4$ & -0.130346+0.638145$R_4$ & 0.1299318$R_4$+1.38115 &
0.569760$R_4$+1.45066\\
8 & 0.245912+0.00135179$R_4$ & -4.61451+0.214571$R_4$ & 0.00135179$R_4$+0.216500
& 0.214531$R_4$+0.0468084\\
9 & -0.342477-0.0104297$R_4$ & -7.59305+0.136738$R_4$ & -0.0104297$R_4$+0.257228 &
0.130480$R_4$+0.0677118\\
10 & -0.880095-0.0108500$R_4$ & -11.5003+0.099648$R_4$ & -0.0184997$R_4$+0.542982 &
0.0842074$R_4$+0.317008\\
11 & -1.65139 - 0.00886659$R_4$ & -17.0050+0.0771335$R_4$ & -0.00886659$R_4$ +
0.894061 & 0.0545640$R_4$ + 0.935218\\
12 & -2.93401-0.00655509$R_4$ & -25.2430+0.0620604$R_4$ & -0.00655509$R_4$+1.23265
& 0.0347475$R_4$+1.97982\\
13 & -5.18889-0.00448899$R_4$ & -38.6692+0.0514389$R_4$ & -0.00448899$R_4$+1.52540
& 0.0212985$R_4$+3.42939\\
14 & -9.53837-0.00279507$R_4$ & -63.6700+0.0437023$R_4$ & -0.00279507$R_4$+1.76163
& 0.0121180$R_4$+5.22737\\
15 & -20.0584-0.00145806$R_4$ & -123.626+0.0379240$R_4$ & -0.00145806$R_4$+1.94165
& 0.00584673$R_4$+7.30916\\
16 & -73.4295-0.000423006$R_4$ & -428.807+0.0335175$R_4$ & -0.000423006$R_4$ +
2.07047 & 0.00158053$R_4$+9.61460
\end{tabular}
\caption{Coefficients $a_{1,2}$ and $b_{1,2}$ of $\beta^{[2|2]}$, the $[2|2]$
Pad\'e-approximant (2.13) to the QCD $\beta$ function for $n_f = 0 - 16$.  The 
coefficients are all linear in $R_4$, the (presently-) unknown five-loop term in (2.10).} 
\label{tab2}
\end{center}
\end{table} 

\begin{table}[htb]
\small
\begin{center}
\begin{tabular}{ccccc}
$n_f$ & $a_1$ & $b_1$ & $b_2$ & $b_3$\\
\hline\\
0 & 9.56239-0.0611053$R_4$ & 7.24421-0.0611053$R_4$ & -24.9099+0.141653$R_4$ &
-42.5901+0.167582$R_4$\\
1 & 8.51104-0.0702708$R_4$ & 6.34975-0.0702708$R_4$ & -20.7090+0.151876$R_4$ &
-33.9265+0.162617$R_4$\\
2 & 7.25658-0.0810329$R_4$ & 5.27382-0.0810329$R_4$ & -16.2327+0.160669$R_4$ &
-25.7265+0.149477$R_4$\\
3 & 5.80845-0.0933552$R_4$ & 4.03067-0.0933552$R_4$ & -11.6367+0.165965$R_4$ &
-18.3242+0.122349$R_4$\\
4 & 4.24716-0.107164$R_4$ & 2.70716-0.107164$R_4$ & -7.21667+0.165032$R_4$ &
-12.2028+0.07244688$R_4$\\
5 & 2.76704-0.123131$R_4$ & 1.50617-0.123131$R_4$ & -3.37387+0.155253$R_4$ &
-7.80319-0.0141605$R_4$\\
6 & 1.72453-0.145814$R_4$ & 0.795958-0.145814$R_4$ & -0.448926+0.135399$R_4$ &
-4.87066-0.168040$R_4$\\
7 & 1.97486-0.200029$R_4$ & 1.44855-0.200029$R_4$ & 1.54554+0.105278$R_4$ &
0.105614-0.517062$R_4$\\
8 & 17.0270-0.778954$R_4$ & 16.9976-0.7789537917$R_4$ & 4.16776+0.0229104$R_4$ &
78.2076-3.63659$R_4$\\
9 & -8.58112+0.137934$R_4$ & -7.98112+0.137934$R_4$ & 2.71758+0.0827604$R_4$ &
-60.2514+1.08502$R_4$\\
10 & -6.27351+0.0358829$R_4$ & -4.85044+0.0358829$R_4$ & 4.14206+0.0510641$R_4$ &
-54.12482+0.468980$R_4$\\
11 & -6.36697+0.0125229$R_4$ & -3.82151+0.0125229$R_4$ & 5.93697+0.0318765$R_4$ &
-61.1352+0.277305$R_4$\\
12 & -7.44146+0.00452652$R_4$ & -3.27479+0.00452652$R_4$ & 8.44183+0.0188605$R_4$
& -72.6300+0.178562$R_4$\\
13 & -9.70448+0.00151777$R_4$ & -2.99019+0.00151777$R_4$ & 11.7796+0.010191$R_4$
& -87.7855+0.116775$R_4$\\
14 & -14.2164+0.000415849$R_4$ & -2.91637+0.000415849$R_4$ &
16.0360+0.00469909$R_4$ & -107.043+0.0734726$R_4$\\
15 & -25.0410+7.04348$\cdot$10$^{-5}$$R_4$ & -3.04098+7.04348$\cdot$10$^{-5}$$R_4$
& 21.3172+0.00154956$R_4$ & -131.384+0.0403041$R_4$\\
16 & -78.8684+2.12019$\cdot$10$^{-6}$$R_4$ & -3.36839+2.12019$\cdot$10$^{-6}$$R_4$
& 27.7873+0.000160074$R_4$ & -162.269+0.0126837$R_4$
\end{tabular}
\caption{Coefficients $a_1$ and $b_{1,2,3}$ of $\beta^{[1|3]}$, the $[1|3]$ Pad\'e-approximant
(2.14)
to the QCD $\beta$-function for $n_f = 0-16$.  $R_4$ is the unknown five-loop contribution
to the $\beta$-function.}
\label{tab3}
\end{center}
\end{table}

\begin{table}[htb]
\small
\begin{center}
\begin{tabular}{ccccc}
$n_f$ & $a_1$ & $a_2$ & $a_3$ & $b_1$\\
\hline\\
0 & 2.31818-0.0240742$R_4$ & 8.11648-0.0558083$R_4$ & 41.5383-0.195397$R_4$ &
-0.0240742$R_4$\\
1 & 2.16129-0.0291294$R_4$ & 6.98533-0.0629572$R_4$ & 34.3295-0.203479$R_4$ &
-0.0291294$R_4$\\
2 & 1.98276-0.0364292$R_4$ & 5.77598-0.0722302$R_4$ & 27.4505-0.210414$R_4$ &
-0.0364292$R_4$\\
3 & 1.77778-0.0476412$R_4$ & 4.47106-0.0846954$R_4$ & 20.9902-0.213007$R_4$ &
-0.0476412$R_4$\\
4 & 1.54000-0.0663747$R_4$ & 3.04764-0.102217$R_4$ & 15.0650-0.202286$R_4$ &
-0.0663747$R_4$\\
5 & 1.26087-0.101668$R_4$ & 1.47479-0.128190$R_4$ & 9.83592-0.149939$R_4$ &
-0.101668$R_4$\\
6 & 0.928571-0.181209$R_4$ & -0.290179-0.168265$R_4$ & 5.51849+0.0525830$R_4$ &
-0.181209$R_4$\\
7 & 0.526316-0.412526$R_4$ & -2.30793-0.217119$R_4$ & 2.42409+0.952081$R_4$ &
-0.412526$R_4$\\
8 & 0.0294118-0.990906$R_4$ & -4.66769-0.0291443$R_4$ & 1.00918+4.62524$R_4$ &
-0.990906$R_4$\\
9 & -0.600000-0.506673$R_4$ & -7.50625+0.304004$R_4$ & 1.97366+3.80322$R_4$ &
-0.506673$R_4$\\
10 & -1.42308-0.155083$R_4$ & -11.0446+0.220695$R_4$ & 6.44815 + 1.71283$R_4$ &
-0.155083$R_4$\\
11 & -2.54545-0.0610294$R_4$ & -15.6645+0.155348$R_4$ & 16.3855+0.955993$R_4$ &
-0.0610294$R_4$\\
12 & -4.16667-0.0281892$R_4$ & -22.0868+0.117455$R_4$ & 35.4746+0.622609$R_4$ &
-0.0281892$R_4$\\
13 & -6.71429-0.0139626$R_4$ & -31.8566+0.0937489$R_4$ & 71.6199+0.444801$R_4$ &
-0.0139626$R_4$\\
14 & -11.3000-0.00687884$R_4$ & -48.9910+0.077730$R_4$ & 145.373+0.337001$R_4$ &
-0.00687884$R_4$\\
15 & -22.0000-0.00301122$R_4$ & -88.2188 + 0.0662468$R_4$ & 332.091+0.265646$R_4$ &
-0.00301122$R_4$\\
16 & -75.500-0.000763368$R_4$ & -282.101+0.0576343$R_4$ & 1309.98 + 0.215347$R_4$ &
-0.000763368$R_4$
\end{tabular}
\caption{Coefficients $a_{1,2,3}$ and $b_1$ of $\beta^{[3|1]}$, the $[3|1]$ Pad\'e-approximant
(2.15) to the QCD $\beta$-function for $n_f=0-16$.  $R_4$ is the unknown five-loop contribution
to the $\beta$-function.}
\label{tab4}
\end{center}
\end{table}

\section{Gluodynamics}

In this section we consider conventional $(N_c = 3)$ QCD with $n_f = 0$.  
The $N_c \rightarrow \infty$ case is considered separately in an Appendix. This
``gluodynamic'' limit is of particular interest as a possible projection (without gluinos)
of supersymmetric QCD in the absence of fundamental-representation matter fields ($n_f=0$ SQCD), a
theory for which the $\beta$-function is known to all orders of
perturbation theory \cite{NSV}.  The $n_f=0$ SQCD $\beta$-function (2.6)
does not exhibit an infrared-stable fixed point;  rather, it exhibits
the dynamics of Fig. 2 in which a $\beta$-function pole serves as an
infrared-attractor, both for a weak asymptotically-free phase, as well as
for a strong phase of the now double-valued couplant \cite{KS}.  Such
dynamics differ fundamentally from those of Fig. 1 anticipated from an
infrared-stable fixed point, which has been argued elsewhere \cite{ACM}
to occur for QCD even in the $n_f = 0$ gluodynamic limit.  
Pad\'e-approximant estimates of higher order terms in QCD $\beta$-functions 
\cite{EJJ} and correlators \cite{CE} have in fact proven to be most accurate in the $n_f=0$
case in which ``quadratic-Casimir'' effects are minimal \cite{EJJ}.
Thus, the methodological machinery of the previous section may be particularly well-suited
to shed insight on whether the dynamics of Fig. 1 or Fig. 2 characterize QCD's
gluodynamic limit.

For $n_f=0$, the four-loop $\overline{MS}$ QCD $\beta$-function is 
\cite{RVL}

\renewcommand{\theequation}{3.\arabic{equation}}
\setcounter{equation}{0}

\begin{equation}
\beta^{(4)}(x)=-\frac{11}{4} x^2 \left[ 1+2.31818x + 8.11648 x^2 +
41.5383 x^3 \right],
\end{equation}
as is evident from substitution of the $n_f=0$ Table \ref{tab1} entries into (2.10).  This $\beta$-function
is sufficient to determine the $[1|2]$- and $[2|1]$-approximant  $\beta$-functions via (2.12) and (2.13):
\begin{equation}
\beta^{[2|1]} (x) = - \frac{11}{4} x^2 \left[ \frac{1-2.7996 x - 3.7475 x^2}{1-5.1178 x} \right],
\end{equation}
\begin{equation}
\beta^{[1|2]} (x) = - \frac{11}{4} x^2 \left[ \frac{1-5.9672x}{1-8.2854x + 11.091x^2} \right],
\end{equation}
In both (3.2) and (3.3), the first positive denominator zero $(x_d)$ precedes
the first positive numerator zero $(x_n)$.  We find from (3.2) that
$x_d=0.195 < x_n = 0.264$, and from (3.3) that $x_d = 0.151 < x_n =
0.168$.  As discussed in Section 2, this $0<x_d<x_n$ ordering of
positive zeros is consistent with dynamics in which $x_d$ serves as in
infrared attractor for both a strong and a weak asymptotically-free
phase (Fig. 2).  The first positive numerator zero $x_n$, if taken
seriously, necessarily corresponds to an ultraviolet-stable fixed point
because of the $\beta$-function sign-change occurring as $x$ passes
through $x_d$.

To test the stability of these conclusions against higher-than-four loop
corrections, we add an arbitrary ``five-loop'' correction to (3.1):
\begin{equation}
\beta^{(5)} (x) = \beta^{(4)}(x) - \frac{11}{4} x^2 [R_4 x^4].
\end{equation}
The five-loop $\beta$-function (3.4) determines $[2|2], [1|3]$, and
$[3|1]$ Pad\'e-approximant $\beta$-functions via (2.14), (2.15) and
(2.16).  These can be read off Tables \ref{tab2}, \ref{tab3} and \ref{tab4}; we list them explicitly
here to facilitate the analysis which follows:
\begin{equation}
\beta^{[2|2]} (x) =
-\frac{11}{4} x^2 \left[ \frac{1+(13.403-0.076216 R_4)x
- (22.915-0.090166 R_4)x^2}{1+(11.084-0.076216 R_4)x 
+ (56.728-0.26685 R_4)x^2}\right]
\end{equation}

\begin{equation}
\beta^{[1|3]}(x)=
-\frac{11}{4} x^2 \left[ \frac{1+(9.5624-0.061105 R_4)x}
{1+(7.2442-0.061105 R_4)x - (24.910-0.14165R_4)x^2 
- (42.590-0.16758R_4)x^3} \right]
\end{equation}

\begin{equation}
\beta^{[3|1]} (x) = -\frac{11}{4} x^2 \left[ \frac{1+(2.31818-0.024074R_4)x + 
(8.1165-0.055808R_4) x^2 + (41.538-0.19540R_4)x^3}{1-0.024074 R_4 x} \right]
\end{equation}

Figure 3  exhibits a plot of $x_n$, the first positive zero of (3.5), and $x_d$ (the 
first positive denominator zero of (3.5),
as a function of the independent variable $R_4$.  Such positive zeros are seen to occur 
for all $R_4$.  Moreover, the first positive denominator zero
is seen to precede the first positive numerator zero over the {\it entire range} of $R_4$.  
This last result confirms that the Fig. 2 dynamics predicted from $[1|2]$ and $[2|1]$ 
approximants appear to be stable against 5-loop corrections of {\it arbitrary} magnitude.

These results are corroborated by $\beta^{[1|3]}$ and $\beta^{[3|1]}$.  Figure 4 plots
the first positive numerator and denominator zeros of (3.6) against
$R_4$ for $\beta^{[1|3]}$, as given by (3.6).  A positive numerator zero
$(x_n)$ exists only if $R_4 > 150$, whereas at least one positive
denominator zero $(x_d)$ occurs for all $R_4$.  Once again, however,
$x_d$ precedes $x_n$ over the entire range of $R_4$, suggesting that
$x_d$ exists as an infrared-attractor for all values of $R_4$.
Moreover, the ordering $0 < x_d < x_n$, over all values of $R_4$ for
which $x_n$ exists, precludes any identification of $x_n$ with an
infrared-stable fixed point.

Figure 5 plots $x_n$ and $x_d$ against $R_4$ for $\beta^{[3|1]}$, as
given in (3.7).  This case is perhaps the most interesting of all, as a
positive denominator root $x_d$ is possible only if $R_4 > 0$, as is
evident from the denominator of (3.7).  Figure 5 shows that $x_d$ continues to precede $x_n$
for all positive values of $R_4$.  Moreover, when $R_4$ is negative (and
a positive pole $x_d$ is no longer possible), one sees from (3.7) that
the numerator polynomial coefficients are all positive-definite,
precluding any possibility of positive {\it numerator} roots.  We
thus see that an infrared-stable fixed point for $\beta^{[3|1]}$ is
unattainable, even for the $R_4 < 0$ region for which no denominator
zero occurs at all.

Thus, there does not exist {\it any} valid $N+M=4$ approximant $(M>0)$
to the $n_f = 0$ QCD $\beta$-function which supports an infrared-stable
fixed point, regardless of the magnitude of the presently unknown 5-loop
term entering such approximants. \footnote{$\beta^{[0|4]}$ cannot have a
nonzero fixed point, as this approximant has no positive numerator zeros
at all.  The $M=0$ case [$\beta^{[4|0]}$], corresponding to the truncated 
series itself,
exhibits a positive zero identifiable with an infrared fixed point only
if $R_4 < 0$, in which case the negative five-loop term must be equal in 
magnitude to the sum of the preceding (positive) one-through-four-loop terms.}  
Moreover, the Fig. 2 dynamics following from a positive
$\beta$-function pole preceding any $\beta$-function zeros appear to be
upheld not only by $[2|1]$- and $[1|2]$-approximant $\beta$-functions,
but also by $[2|2]$-, $[1|3]$- and (when $R_4 > 0$) $[3|1]$-approximant
$\beta$-functions as well.

One can also utilize asymptotic Pad\'e approximant methods to estimate the
magnitude of the infrared attractor (Fig. 2), corresponding to the first
positive denominator zero of all the approximants considered so far.  An
asymptotic error formula \cite{EJJ} enables one to obtain an asymptotic
Pad\'e-approximant prediction (APAP) of $R_4$ from the three preceding
terms $\{R_1, R_2, R_3\}$ in the $\beta$-function series \cite{ESC}:
\footnote{This formula summarizes the content of the algorithm 
presented in Sections 2 and 5 of ref. \protect\cite{EJJ}}
\begin{equation}
R_4^{APAP} = \frac{R_3^2}{R_2} \frac{[R_2^3+R_1R_2R_3 -
2R_1^3R_3]}{[2R_2^3-R_1^3R_3 - R_1^2R_2^2]}.
\end{equation}
Utilizing the values $\{R_1,R_2,R_3\}$ listed in Table \ref{tab1} for $n_f=0$,
one finds from (3.8) that $R_4^{APAP} = 302.2$.  If one inserts this
value of $R_4$ into the $[2|2]$-, $[1|3]$- and $[3|1]$-approximant
$\beta$-functions (3.5), (3.6) and (3.7), one finds reasonable agreement
from all three $\beta$-functions as to the location of the first
positive denominator zero, {\it i.e.} the infrared-attractor of Fig. 2.
This zero is seen to occur at $x_d = 0.11$ for $\beta^{[2|2]}$ and
$\beta^{[1|3]}$, and at $x_d = 0.14$ for $\beta^{[3|1]}$, corresponding
to values of $\alpha_s$ between 0.35 and 0.44.
A similar $R_4$-estimate ($R_4=270$) can be obtained for the $n_f=0$ case via the weighted 
asymptotic Pad\'e approximant procedure (WAPAP) delineated in Section 5 
of  ref. \cite{EJJ}.
This procedure expresses the coefficient $\beta_4=\beta_0R_4$ as a degree-4 polynomial in $n_f$,
as would be obtained from an explicit perturbative calculation.
\footnote{Coefficients $\beta_k$ listed in \protect\cite{EJJ} must be divided
by $4^{k+1}$ to correspond to our normalization of $\beta$-function coefficients:
our $\beta(x)=dx/d(\log\mu^2)$ with $x=\alpha/\pi$.}  We reiterate that the $n_f=0$ 
case minimizes unknown quadratic Casimir effects and avoids entirely the large uncertainties
associated with the cancellation of large $n_f$-dependent terms within $\beta_4$
(as evident from eq. (5.5) of \cite{EJJ}) that characterize $n_f>0$ estimates based on 
APAP and WAPAP methods.

We conclude this section by reiterating that the ordering $0<x_d<x_n$ 
of positive denominator and numerator zeros of $n_f = 0$ Pad\'e-approximant $\beta$-functions
suggests the existence of a double-valued couplant (Fig. 2), as already seen in SUSY
gluodynamics \cite{KS}.  Such a scenario is seen to decouple the infrared region $(\mu < \mu_c)$ from the domain
of purely-perturbative QCD, the domain of (real) $\alpha_s$.  Such a scenario is perhaps also
indicative of an additional phase distinguished by strong coupling dynamics at short
distances \footnote{It should
be noted here that the presence of an infrared-stable fixed point $\alpha^*$ 
also implies a possible strong
phase of the couplant \protect\cite{MY}, 
in addition to the asymptotically-free phase exhibited in Fig. 1.  If
$\alpha(\mu) > \alpha^* [\mu > 0]$, then $\alpha$ approaches $\alpha^*$
{\it from above} as $\mu$ approaches zero from above.} Taking the Pad\'e-approximants
of the $\beta$-function seriously, one concludes that there is an ultraviolet-stable
fixed point in that phase.  Recall that in SUSY gluodynamics, such a fixed point is at
$x = \infty$ \cite{KS}. Of course, it is impossible to assert that this ultraviolet-stable
fixed point survives in the exact $\beta$-function of (non-SUSY) gluodynamics.  For example,
it may be replaced by an ultraviolet Landau pole.

Regardless of these considerations, the picture with an infrared attractor seems plausible and 
self-consistent.  In such dynamics, both phases may share common infrared properties \cite{KS}.
The presence of two phases has implications meriting further exploration.  
Such dynamics are shown in the Appendix to characterize QCD in the $N_c \rightarrow \infty$ limit 
for all but the aforementioned $[3|1]$-case with $R_4$ negative.  Indeed, such dynamics may 
prove pertinent to the unexpected agreement between the glueball mass spectra obtained 
via lattice methods \cite{MJT} and those obtained via supergravity wave equations in a black 
hole geometry \cite{CC} following from conjectured duality to large-N$_c$ gauge theories \cite{JM},
an agreement obtained despite the large bare coupling constant necessarily utilized in the latter approach.

In the section which follows, we extend the analysis of the QCD $\beta$-function to nonzero
$n_f$ values.  We specifically seek to address whether Pad\'e-methods indicate a flavour-threshold for 
infrared-stable fixed points.  However, we also seek insight as to whether
there is any evidence for a strong phase of QCD when $n_f > 0$, as such a phase could (conceivably) 
provide a dynamical mechanism for electroweak symmetry breaking.

\section{QCD with Fermions}

In this Section, we repeat the analysis of the previous section with
$n_f \leq 16$ fermion flavours.  The $n_f = 16$ case is the maximum
number of flavours consistent with asymptotic freedom: a $\beta$-function
whose sign is negative as $x \rightarrow 0^+$.

\subsection{\it Four-Loop-Level Results}

We first consider the $\beta$-function (2.12) incorporating a $[2|1]$
approximant to describe post-one-loop behaviour.  This $\beta$-function
is fully determined by the known two-, three-, and four-loop terms
$\{R_1, R_2, R_3\}$ tabulated in Table \ref{tab1} for $n_f = \{1,2,...,16\}$.
The values of the coefficients $\{a_1, a_2, b_1\}$ characterizing
$\beta^{[2|1]}$ in (2.12a) are tabulated in Table \ref{tab5}.  Also tabulated in the
table are values for $x_n$, the first positive zero of the numerator
$1+a_1 x + a_2 x^2$, as well as $x_d$, the zero of the denominator
$1+b_1 x$.  Blank entries for $x_n, x_d$ correspond to cases where no
positive zero exists.  We see from the table that a positive denominator
zero precedes the first positive numerator zero for $n_f \leq 5$,
in which case that denominator zero serves as an infrared attractor for
both a strong and an asymptotically-free ultraviolet phase of the QCD
couplant.  The denominator zero becomes negative for $n_f \geq 6$;
nevertheless, an infrared-stable fixed point (associated with a positive
numerator zero) does not occur until $n_f = 7$,  as no positive
numerator zero exists for $n_f = 6$. The value of $x_n = \alpha_s / \pi$
associated with this infrared-stable fixed point decreases as $n_f$
increases, as anticipated in other work \cite{ATW,MY}.  If we regard
$\alpha_{cr} = \pi / 4$ ($x_n = 1/4$) as the threshold value for
chiral symmetry breaking \cite{FGM}, we see that the conformal window
for QCD is predicted to begin at $n_f = 11$.  For $n_f = \{ 7-10 \}$, the
infrared-stable fixed point $x_n$ is not expected to govern infrared dynamics.
Chiral-symmetry breaking should occur before the couplant reaches $x_n$, and the
(now-massive) fermions are expected to decouple from further infrared evolution
of $\alpha_s$ \cite{ATW,MY}. 

\begin{table}[htb]
\small
\begin{center}
\begin{tabular}{cccccc}
$n_f$ & $a_1$ & $a_2$ & $b_1$ & $x_n$ & $x_d$\\
\hline\\
0 & -2.79959 & -3.74745 & -5.11777 & 0.26394 & 0.19540\\
1 & -2.75323 & -3.63638 & -4.91452 & 0.26820 & 0.20350\\
2 & -2.76977 & -3.64714 & -4.75253 & 0.26710 & 0.21041\\
3 & -2.91691 & -3.87504 & -4.69468 & 0.25586 & 0.21301\\
4 & -3.40349 & -4.56534 & -4.94349 & 0.22557 & 0.20229\\
5 & -5.40850 & -6.93442 & -6.66937 & 0.15435 & 0.14994\\
6 & 19.9461 & 17.3690 & 19.01757 & --- & ---\\
7 & 1.57664 & -1.75513 & 1.05033 & 1.3275 & ---\\
8 & 0.245617 & -4.66133 & 0.216205 & 0.49027 & ---\\
9 & -0.337065 & -7.66401 & 0.262935 & 0.33990 & ---\\
10 & -0.839249 & -11.8754 & 0.583828 & 0.25699 & ---\\
11 & -1.49942 & -18.3271 & 1.04603 & 0.19624 & ---\\
12 & -2.56052 & -28.7791 & 1.60615 & 0.14716 & ---\\
13 & -4.46609 & -46.9517 & 2.24819 & 0.10593 & ---\\
14 & -8.33265 & -82.5220 & 2.96735 & 0.070620 & ---\\
15 & -18.2356 & -171.036 & 3.76441 & 0.039903 & ---\\
16 & -70.8563 & -632.698 & 4.64367 & 0.012678 & ---
\end{tabular}
\caption{The coefficients $a_1$, $a_2$ and $b_1$ in $\beta^{[2|1]}$, as given
in Eq. (2.12) for $n_f = 0 - 16$.  The column $x_n$ tabulates the first positive
zero of the numerator.  The column $x_d$ tabulates the denominator zero, if positive.  For 
those values of $n_f$ for which real positive numerator (denominator) zeros do not exist,
the corresponding entries for $x_n \; (x_d)$ are left blank. The table shows $0 < x_d < x_n$ for $n_f \leq 5$,
corresponding to Fig. 2 type dynamics for this range of $n_f$.  When $n_f \geq 7$, $x_n$ corresponds
to an infrared-stable fixed point that decreases as $n_f$ increases.}
\label{tab5}
\end{center}
\end{table}

Qualitatively similar conclusions are obtained from the analysis of
$\beta^{[1|2]}$ [eq. (2.13)], although the corresponding values of 
$n_f$-thresholds
for various infrared properties differ somewhat from those of the
$[2|1]$-approximant case.  In Table \ref{tab6} the values of the constants
$\{a_1, b_1, b_2\}$ characterizing $\beta^{[1|2]}$ in (2.13a) are tabulated
using the known values for $\{R_1, R_2, R_3\}$ listed in Table \ref{tab1}.  When
positive, the zero of the numerator $1+a_1 x$ fails to precede positive
zeros of the denominator $1+b_1 x + b_2 x^2$ until $n_f = 9$ (positive
denominator zeros cease occurring after $n_f = 6$, but the numerator
zero is negative when $5 \leq n_f \leq 8$).  Consequently, $n_f = 9$ is
the flavour-threshold for identification of $x_n$ as an infrared-stable
fixed point.  As before, this fixed point decreases with increasing
$n_f$.  Its magnitude does not fall below the $x_n =1/4$ threshold 
for chiral-symmetry breakdown until $n_f = 12$, corresponding to a
previous prediction \cite{ATW} of the threshold for QCD's conformal
window, consistent with the qualitative picture presented in
ref. \cite{MY}. Specific predictions for $x_n$ from Tables \ref{tab5} and \ref{tab6} agree
quantitatively only within this $12 \leq n_f \leq 16$ window.  
Although the intermediate range of $n_f$ for which $x_n > 1/4$ differs 
between the two
approximants ($7 \leq n_f \leq 10$ for $\beta^{[2|1]}$;  $9 \leq n_f
\leq 11$ for $\beta^{[1|2]}$), it is nevertheless significant that such
a range exists for both cases.  In Table \ref{tab6}, a positive denominator zero
is seen to precede any positive numerator zeros when $n_f \leq 6$.  This
behaviour corresponds to the infrared dynamics suggested by Figure 2.
Once again, qualitative agreement is seen to occur between the $[2|1]$-
and $[1|2]$-approximant $\beta$-functions insofar as both $\beta$-
functions predict dynamics governed by a $\beta$-function pole $x_d$ for
$n_f$ less than some threshold value ($n_f \leq 5$ for $\beta^{[2|1]}$;
$n_f \leq 6$ for $\beta^{[1|2]}$).

\begin{table}[htb]
\small
\begin{center}
\begin{tabular}{cccccc}
$n_f$ & $a_1$ & $b_1$ & $b_2$ & $x_n$ & $x_d$\\
\hline\\
0 & -5.96723 & -8.28541 & 11.0906 & 0.16758 & 0.15136\\
1 & -6.14941 & -8.31070 & 10.9765 & 0.16262 & 0.15007\\
2 & -6.68998 & -8.67274 & 11.4200 & 0.14948 & 0.14177\\
3 & -8.17337 & -9.95114 & 13.2199 & 0.12235 & 0.11944\\
4 & -13.8032 & -15.3432 & 20.5809 & 0.072447 & 0.072160\\
5 & 70.6188 & 69.3580 & -88.9261 & --- & 0.79411\\
6 & 5.95098 & 5.02241 & -4.37349 & --- & 1.32141\\
7 & 1.93401 & 1.40769 & 1.56704 & --- & ---\\
8 & 0.274983 & 0.245571 & 4.66047 & --- & ---\\
9 & -0.921639 & -0.321639 & 7.31327 & 1.0850 & ---\\
10 & -2.13228 & -0.709208 & 10.0353 & 0.46898 & ---\\
11 & -3.60614 & -1.06069 & 12.9645 & 0.27730 & ---\\
12 & -5.60030 & -1.43363 & 16.1133 & 0.17856 & ---\\
13 & -8.56349 & -1.84921 & 19.4406 & 0.11677 & ---\\
14 & -13.6105 & -2.31052 & 22.8821 & 0.073472 & ---\\
15 & -24.8114 & -2.81137 & 26.3685 & 0.040304 & ---\\
16 & -78.8413 & -3.34126 & 29.8352 & 0.012684 & ---
\end{tabular}
\caption{The coefficients $a_1$, $b_1$ and $b_2$ in $\beta^{[1|2]}$ as given in Eq.
(2.13) for $n_f = 0-16$.  The column $x_d$ tabulates the first positive denominator zero.  The 
column $x_n$ tabulates the numerator zero, if positive.  For those values of $n_f$ for which real positive
denominator (numerator) zeros do not exist, the corresponding entries for $x_d \; (x_n)$
are left blank.  The positive denominator zero $x_d$ occurs before any positive numerator
zeros for $n_f \leq 6$, suggesting Fig. 2 type dynamics. When $n_f \geq 9$, $x_n$ corresponds
to an infrared-stable fixed point that decreases as $n_f$ increases}
\label{tab6}
\end{center}
\end{table}

\subsection{\it Five-Loop Level Results}

It is of interest to examine the stability of the qualitative results
described above against five-loop corrections to the $\overline{MS}$ $\beta$-function,
corrections which do not enter our determination of $\beta^{[2|1]}$ and
$\beta^{[1|2]}$.  As noted in Section 2,  Pad\'e-coefficients within
$\beta^{[2|2]}$, $\beta^{[1|3]}$ and $\beta^{[3|1]}$ are all seen to be
linear in the five-loop $\beta$-function correction $\beta_4$
($\beta_4/\beta_0 \equiv R_4$).  These coefficients, as defined by
equations (2.14a), (2.15a), and (2.16a), are respectively tabulated in
Tables \ref{tab2}, \ref{tab3} and \ref{tab4}. In 
Table \ref{tab7} we have tabulated the domain of $R_4$ for
which the first positive numerator zero of the Pad\'e-approximant
$\beta$-function precedes any positive denominator zero.  Such a
numerator zero implies a couplant with Figure 1 type dynamics, in which
the numerator zero is an infrared-stable fixed point.  We see from Table
\ref{tab7} that such dynamics do not occur at all {\it regardless of $R_4$}
unless $n_f \geq 6$.  An infrared-stable fixed point cannot occur for
$\beta^{[2|2]}$ until $n_f = 7$ (and then only for $R_4 < 0$), nor can it
occur for $\beta^{[1|3]}$ until $n_f = 9$.  As $n_f$ increases, the
domain of $R_4$ for which Figure 1 type dynamics become possible is seen
to broaden for all three Pad\'e-approximant $\beta$-functions.  The
overall picture that emerges is quite similar to that anticipated from
Tables 5 and 6.  In every case, dynamics governed by an infrared-stable
fixed point (Fig. 1) do not occur below a threshold value of $n_f$, a
threshold at or above $n_f = 6$.

Table 8 tabulates the range of $R_4$ for which a positive pole of
$\beta^{[2|2]}$, $\beta^{[1|3]}$ and $\beta^{[3|1]}$ exists and precedes
any positive numerator zeros, corresponding to the dynamics
schematically presented in Figure 2.  For $n_f \leq 5$, $\beta^{[2|2]}$
and $\beta^{[1|3]}$ are seen to exhibit such dynamics regardless of the
magnitude of $R_4$, the five-loop contribution to the $\beta$-function.
$\beta^{[3|1]}$ exhibits such Figure 2 type dynamics only if $R_4$ is
positive, as the denominator $1+b_1 x$ of eq. (2.16a) has a positive zero
only if $R_4 > 0$ $[b_1=-R_4/R_3]$.  Such dynamics, however, become
impossible for $\beta^{[1|3]}$ and unlikely for $\beta^{[2|2]}$ and
$\beta^{[3|1]}$ once $n_f$ gets sufficiently large, as is apparent in
Table 8 from the steadily increasing lower bound on $R_4$ for such
dynamics to occur within $\beta^{[2|2]}$ and $\beta^{[3|1]}$.

This large $n_f$ behaviour is illustrated for $n_f = 13$ by plots of the
$R_4$-dependence of the first positive numerator and denominator zero of
$\beta^{[2|2]}$ (Fig. 6), $\beta^{[1|3]}$ (Fig. 7) and $\beta^{[3|1]}$
(Fig. 8).  The infrared-stable fixed point
associated with the numerator zero in $\beta^{[2|2]}$ and $\beta^{[3|1]}$
is seen to be less than 1/4
(the assumed threshold for chiral-symmetry breakdown \cite{FGM}) over
the full domain of $R_4$ indicated in Table \ref{tab7} for infrared-stable fixed
point dynamics, a result is clearly suggestive of $n_f = 13$ being
within the conformal window of QCD.  The numerator zero in $\beta^{[1|3]}$ is also
below 1/4, as evident from Fig. 7, until the immediate neighbourhood of its 
singularity at $R_4 = 6394$.

\begin{table}[htb]
\small
\begin{center}
\begin{tabular}{cccc}
$n_f$ & $[2|2]$ & $[1|3]$ & $[3|1]$\\
\hline\\
0 & No $R_4$& No$R_4$ & No$R_4$\\  
1 & No $R_4$& No$R_4$ & No$R_4$\\
2 & No $R_4$& No$R_4$ & No$R_4$\\
3 & No $R_4$& No$R_4$ & No$R_4$\\
4 & No $R_4$& No$R_4$ & No$R_4$\\
5 & No $R_4$& No$R_4$ & No$R_4$\\
6 & No $R_4$& No$R_4$ & $R_4 < -105$\\
7 & $R_4 < 0.21$ & No$R_4$ & $R_4 < -2.1$\\
8 & $R_4 < 21.6$ & No$R_4$ & $R_4 < 1.2$\\ 
9 & $R_4 < 57.2$ & $R_4 < 62.3$ & $R_4 < 3.6$\\
10 & $R_4 < 128.4$ & $R_4 < 174.9$ & $R_4 < 12.9$\\
11 & $R_4 < 274.6$ & $R_4 < 508.5$ & $R_4 < 38.4$ \\
12 & $R_4 < 594.9$ & $R_4 < 1644$ & $R_4 < 110$\\ 
13 & $R_4 < 1384$ & $R_4 < 6394$ & $R_4 < 340.3$\\
14 & $R_4 < 3750$ & $R_4 < 34187$ & $R_4 < 1226$\\
15 & $R_4 < 14262$ & $R_4 < 355521$ & $R_4 < 6057$\\ 
16 & $R_4 < 174589$ & $R_4 < 37198800$ & $R_4 < 92883$
\end{tabular}
\caption{The domain of $R_4$ for which an infrared fixed point occurs for
$\beta^{[2|2]}$, $\beta^{[1|3]}$ and $\beta^{[3|1]}$.  The ranges of $R_4$
listed (for a given choice of $n_f$) are those for which a positive
numerator zero
exists and precedes any positive denominator zeros of the Pad\'e-approximant
$\beta$-function.}
\label{tab7}
\end{center}
\end{table}

\begin{table}[htb]
\small
\begin{center}
\begin{tabular}{cccc}
$n_f$ & $[2|2]$ & $[1|3]$ & $[3|1]$\\
\hline\\
0 & All $R_4$ & All $R_4$ & $R_4>0$\\
1 & All $R_4$ & All $R_4$ & $R_4>0$\\
2 & All $R_4$ & All $R_4$ & $R_4>0$\\
3 & All $R_4$ & All $R_4$ & $R_4>0$\\
4 & All $R_4$ & All $R_4$ & $R_4>0$\\
5 & All $R_4$ & All $R_4$ & $R_4>0$\\
6 & $R_4 > -105$ & $R_4 > -35.4$ & $R_4 > 0$\\
7 & No $R_4$ & $R_4 > 0.2$ & $R_4 > 0$\\
8 & No $R_4$ & $R_4 > 21.5$ & $R_4 > 1.2$\\
9 & $R_4 > 4848$ & No $R_4$ & $R_4 > 3.6$\\
10 & $R_4 > 2964$ & No $R_4$ & $R_4 > 12.9$\\
11 & $R_4 > 2990$ & No $R_4$  & $R_4 > 38.4$\\
12 & $R_4 > 3652$ & No $R_4$ & $R_4 > 110$\\
13 & $R_4 > 5020$ & No $R_4$ & $R_4 > 340.3$\\
14 & $R_4 > 7759$ & No $R_4$ & $R_4 > 1226$\\
15 & $R_4 > 14491$ & No $R_4$ & $R_4 > 6057$\\
16 & $R_4 > 174589$ & No $R_4$ & $R_4 > 92883$
\end{tabular}
\caption{The domain of $R_4$ for which a positive denominator zero exists and precedes any positive
numerator zeros of $\beta^{[2|2]}$, $\beta^{[1|3]}$ and $\beta^{[3|1]}$, respectively.
Such a condition corresponds to the Figure 2 scenario for coupling constant evolution, in which
the positive denominator zero is an infrared attractor for both a strong and weak phase of the
couplant.}
\label{tab8}
\end{center}
\end{table}

The decrease of the infrared fixed point with increasing $n_f$ is also
common to all Pad\'e-approximant $\beta$-functions.  This behaviour, as
already seen in Tables 5 and 6 for $\beta^{[2|1]}$ and $\beta^{[1|2]}$,
is illustrated for $\beta^{[2|2]}$ in Fig. 9, in which the magnitudes
of $x_n$ are displayed as functions of $R_4$ for $n_f = \{ 9 - 16 \}$.  
Such results are consistent with the phase structure anticipated
in ref. \cite{MY}, as already noted. It is also worth mentioning that the general picture we obtain,
particularly the need for a critical number of flavours for an infrared-stable fixed point to occur
at all, agrees surprisingly well with a lattice study \cite{Iwasaki}.

\section{QCD's Infrared Boundary}

The case of three flavours is of obvious interest, as Pad\'e-extrapolations
to the infrared region can be compared to the known empirical dynamics
at the onset of the infrared region.  We
know, for example, that evolution of the running coupling constant from
its well-determined value at $\mu = M_z$ \cite{PDG} leads to a prediction
\cite{ESC}

\renewcommand{\theequation}{5.\arabic{equation}}
\setcounter{equation}{0}

\begin{equation}
\alpha_s (n_f = 3; \; \; \mu=1 \; GeV)= 0.48 
\begin{array} {c} +0.09 \\ -0.07
\end{array}
\end{equation}
We also know that QCD {\it as a theory of quarks and gluons} ceases to
exist at momentum scales approaching $\Lambda_{QCD}$, although the interpretation of
$\Lambda_{QCD}$ is subject to redefinition for each successive order of
perturbation theory.

Tables 5-7 show quite clearly that an infrared-stable fixed point for
$n_f = 3$ QCD is unsupported by all Pad\'e-approximant $\beta$-functions
considered here, regardless of the magnitude of $R_4$.  This result 
contradicts the infrared-stable fixed point obtained from the analysis
of a lower-order expression for the $\beta$-function in refs. \cite{MS} and
\cite{ACM}, although the absence of such a fixed point at $n_f = 3$ is supported by more 
recent work \cite{EGM}.  We also note that {\it all} Pad\'e-
approximant $\beta$-functions considered here exhibit Figure 2 type
dynamics, regardless of $R_4$, except for $\beta^{[3|1]}$ when $R_4 <
0$.  In such dynamics, the $\beta$-function pole $x_d$ occurs at
momentum-scale $\mu_c$.  As evident from Fig. 2, the infrared region
$\mu < \mu_c$ is inaccessible to the (real) couplant $x(\mu)$, suggesting
that $\mu_c$ fulfills the infrared cutoff role commonly ascribed to
$\Lambda_{QCD}$, the ``Landau pole'' obtained through use of the truncated
$\beta$-function series.

In Figs. 10-12, we have exhibited the $R_4$ dependence of the $\beta$-
function pole $x_d$ for $\beta^{[2|2]}$, $\beta^{[1|3]}$ and
$\beta^{[3|1]}$ when $n_f = 3$.  The first positive numerator zero is
also displayed in all three figures, and is seen to be larger than $x_d$
for all $R_4$ values considered.  We note from Figs. 10 and 11 the
apparent stability of $x_d$ in $\beta^{[2|2]}$ and $\beta^{[1|3]}$
against changes in $R_4$ when $R_4$ is negative.  Both figures indicate
an infrared-attractor near $x_d = 0.4$ $(\alpha_s \cong 1.3)$, a value 
well-above the anticipated threshold for chiral-symmetry breaking.

Given knowledge of an initial value, one can utilize Pad\'e approximant
$\beta$-functions to estimate the infrared cutoff $\mu_c$.  To
demonstrate this, we assume from the central value of (5.1) that $x(1 \;
GeV) = \alpha_s (1 \; GeV) / \pi = 0.153$.  The equation
\begin{equation}
\mu^2 \frac{dx}{d \mu^2} = \beta^{[N|M]} (x)
\end{equation}
can be inverted to determine $\mu_c$ the value of $\mu$ corresponding to
the first positive pole of $\beta^{[N|M]}$, Figure 2's infrared attractor
$x_d = x(\mu_c)$:
\begin{equation}
\mu_c (GeV) = exp \left[\frac{1}{2} \int_{x(1 GeV)}^{x_d}
\frac{dx'}{\beta^{[N|M]}(x')} \right]
\end{equation}
We have utilized (5.3) to plot predicted values of $\mu_c$ for $[2|2]$; $[1|3]$;
and $[3|1]$-approximant $\beta$-functions against the unknown 5-loop
$\beta$-function coefficient $R_4(\equiv \beta_4/\beta_0)$.  Fig. 13
utilizes $\beta^{[2|2]}$, as determined by the $n_f = 3$ row of Table 2,
to predict $\mu_c$, given $x(1) = 0.153$.
The curve terminates with $\mu_c = 1 \; GeV$ at $R_4 = 128$, since $x_d$
(the infrared attractor) is itself equal to 0.153 at this value of
$R_4$.  What is noteworthy, however, is the stability of $\mu_c$ over
the entire range of negative $R_4$.
Fig. 13 is clearly indicative of
QCD's infrared cutoff (mass gap) occurring not much below the $\rho$ mass:  
$\mu_c \rightarrow 660 \; MeV$ as $R_4 \rightarrow -\infty$.

Thus Fig. 13 is indicative of a lower bound for $\mu_c$ well-above the
phenomenological value for $\Lambda_{QCD}$ when $n_f = 3$.  The bound on
$\mu_c$ obtained from $\beta^{[2|2]}$ remains well above $\Lambda_{QCD}$
even if the estimate for $\alpha_s (1 \; GeV)$ is reduced to the floor
of its empirical range.  Fig. 14
utilizes $\beta^{[2|2]}$ in conjunction with the lower-bound value of 
(5.1) for $\alpha_s (1 \; GeV)$  $\left[ x(1)=\alpha_s (1 \; GeV) / \pi = 0.1305 \right]$.
The figure continues to predict insensitivity to $R_4$ over the entire negative range of 
$R_4$, with a somewhat diminished lower bound on $\mu_c$:  $\mu_c \rightarrow 550 \; MeV$
from above as $R_4 \rightarrow -\infty$.

The behaviour described above is virtually identical to that obtained by utilizing $\beta^{[1|3]}$ within
(5.3).  Figures 13 and 14 display $\mu_c$ as a function of $R_4$, with $x(1) = 0.153$ [Fig 15]
and $x(1) = 0.1305$ [Fig 16].  The expression for $\beta^{[1|3]}$ utilized in the integrand of (5.3)
can be extracted from the $n_f = 3$ row of Table 3.  Both curves terminate at $\mu_c = 1 \; GeV$ at $R_4$ values corresponding
to the infrared attractor $x_d$ being equal to $x(1 \; GeV)$ [$x_d = 0.153$ and $x_d = 0.1305$, respectively].
Both curves also demonstrate the same stability of $\mu_c$ against changes in $R_4$, as well as 
virtually the same lower bounds for $\mu_c$ as
obtained in Figs. 13 and 14 from $\beta^{[2|2]}$.

As noted earlier, an infrared attractor associated with an $n_f = 3$ $\beta$-function pole occurs within
$\beta^{[3|1]}$ only for $R_4 > 0$.  In Fig. 17, we plot the $\beta^{[3|1]}$-prediction
for the value of $\mu_c$, as obtained from (5.3), against positive values of $R_4$.  Using
the central value $x(1) = 0.153$ from (4.1), we see that $\mu_c > 600 MeV$ over the entire 
(positive) range of $R_4$.

For $R_4 < 0$, $\beta^{[3|1]}(x)$ no longer has a pole.  This does not mean, however, that $\alpha_s(\mu)$
has a domain in which $\mu$ can get arbitrarily close to zero.  Rather, when $R_4 < 0$, there will exist 
a Landau pole, {\it i.e.} a minimum value of $\mu$ at which $\alpha_s$ will diverge:
\begin{equation}
\mu_L = exp \left( \frac{1}{2} \int_{x(1 GeV)}^\infty \frac{dx}{\beta^{[3|1]}} \right) 
\end{equation}
Similar dynamics characterize the evolution of $\alpha_s(\mu)$ that
follows from the one-loop $\beta$-function [$\beta(x) = -\beta_0 x^2$],
even though this $\beta$-function itself has neither poles nor non-zero fixed
points.  We have utilized (5.4) in Fig. 18 to find the minimum value of
$\mu$ as a function of $R_4$.  If $x(1 \; GeV) = 0.153$, consistent with
(5.1), we then find that 
$\mu_L$ approaches 530 MeV from above as $R_4 \rightarrow -\infty$.

In {\it every} case we consider, it is clear that the domain of
$\alpha_s(\mu)$ when $n_f = 3$ is bounded from below by hadronic mass scales 
comparable to or
somewhat below the $\rho$-mass.  Pad\'e-approximant $\beta$-
functions appear to decouple the infrared region from $\alpha_s$ at
values of $\mu$ substantially larger than $\Lambda_{QCD}$.

\section{Conclusions}

    Utilizing Pad\'e-summation QCD $\beta$-functions whose Maclaurin expansions reproduce the
known terms of the $\overline{MS}$ $\beta$-function series, we obtain a surprising degree of agreement 
with infrared properties predicted \cite{ATW, MY, EGM} via the 't Hooft renormalization scheme 
\cite{GH} 
in which the $\beta$-function is truncated subsequent to two-loop order.  Within the context of 
$N_c = 3$ QCD, we find clear evidence for a flavour-threshold between $n_f = 6$ and $n_f = 9$ 
for any possibility at all of infrared dynamics governed by an infrared-stable fixed point.  
For $n_f < 6$, no approximant-based á-function (other than the truncated series
itself) is able to yield a positive zero that is not preceded by a positive pole,
regardless of the as-yet-unknown magnitude of the five-loop contribution ($R_4$) to the 
$\overline{MS}$ $\beta$-function series (2.10).  We reiterate that such a zero can be identified as an 
infrared-stable fixed point only if it is {\it not} preceded by a positive $\beta$-function pole. 

     We also find (Section 4) that when $n_f$ exceeds the (approximant-dependent) flavour 
threshold for possible infrared-stable fixed-point dynamics, the magnitude of that fixed 
point is seen to decrease as $n_f$ increases (Tables 5 and 6 and Figure 9). The true conformal window 
of QCD is not expected to begin until the infrared-stable fixed point is sufficiently 
small to preclude chiral-symmetry breaking in the infrared region \cite{ATW, MY}.  In Section 4, we 
corroborate this window's onset at 11-13 flavours, as anticipated from two-loop results 
\cite{ATW, MY}.

     For values of $n_f$ below the (approximant-dependent) threshold for a possible infrared-stable 
fixed point, Pad\'e-summations of the $\overline{MS}$ $\beta$-function are indicative of infrared dynamics 
governed by a $\beta$-function pole, as has already been observed for N=1 SQCD in the absence 
of fundamental-representation matter fields \cite{KS}.  Such a positive pole preceding all 
positive zeros of the $\beta$-function is found to occur for $n_f \leq 5$ for all Pad\'e-approximant 
$\beta$-functions constructed from the $\overline{MS}$ $\beta$-function series, regardless of the unknown five-loop 
term ($R_4$) in that series, except for the $\beta$-function incorporating both a $[3|1]$-approximant 
to higher-loop effects and a negative value of $R_4$.  Infrared dynamics governed by such a 
pole have a number of phenomenologically interesting properties. 
Salient among these is the occurrence of an infrared cut-off $\mu_c$ on the domain of the QCD 
couplant $\alpha_s(\mu)/\pi$ (Figure 2). 
We find (Figs. 13-17) the magnitude of $\mu_c$ to be quite stable against 
changes in $R_4$, and to be bounded from below by values 
larger than $\Lambda_{QCD}$ and comparable to low-lying meson masses 
(500-700 MeV).  Such an infrared boundary occurs even for the $[3|1]$-approximant case with 
$R_4 < 0$, a case for which a positive $\beta$-function pole does not exist. For this case the infrared boundary 
corresponds to a Landau pole at similar hadronic mass scales (Fig. 18).

     Pole-dominated infrared dynamics have also been argued \cite{KS} to imply the existence of 
a strong phase that devolves to the same infrared attractor as the asymptotically-free phase 
of the QCD couplant.  In such dynamics, both phases may share common infrared properties 
\cite{KS}.  As we have noted at the end of Section 3, such dynamics may provide a clue as to why 
lattice results for the glueball spectrum appear to be in agreement with results obtained
in the strong-coupling, large-$N_c$ limit of $SU(N_c)$ via supergravity
wave equations within
a black hole geometry.  

\section*{Acknowledgments}

V.E. and T.G.S. are grateful for support from the Natural Sciences and Engineering Research Council 
of Canada.  V.A.M. is grateful to Koichi Yamawaki for his warm hospitality at 
Nagoya University, where this paper was finished.  The work of V.A.M. was supported by the 
Grant-in-Aid of the Japan Society for the Promotion of Science No. 11695030.

\section*{Appendix:  Gluodynamics in the 't Hooft Limit}

If the product of $N_c$ and $\alpha_s$ is finite and nonzero in the $N_c
\rightarrow \infty$ limit, the $n_f = 0 \; {\overline{MS}} \; SU(N_c)$ $\beta$-
function in this same limit is given by \cite{RVL}

\renewcommand{\theequation}{A.1\alph{equation}}
\setcounter{equation}{0}

\begin{equation}
\mu^2 \frac{d \lambda}{d \mu^2} = - \frac{11}{3} \lambda^2 \left[ 1 +
R_1 \lambda + R_2 \lambda^2 + R_3 \lambda^3 (+ \; R_4 \lambda^4 ...) \right]
\end{equation}
\begin{equation}
\lambda \equiv N_c \alpha_s (\mu)/4 \pi,
\end{equation}
\begin{equation}
R_1 = 34/11,
\end{equation}
\begin{equation}
R_2 = 2857/198,
\end{equation}
\begin{equation}
R_3 = 86.04326.
\end{equation}
The coefficients subsequent to $R_3$ in (A.1a) remain unknown at
present.  However, the known coefficients in (A.1a) are sufficient to
determine Pad\'e-summation functions incorporating $[2|1]$ and $[1|2]$
approximants via (2.12) and (2.13):

\renewcommand{\theequation}{A.\arabic{equation}}
\setcounter{equation}{1}

\begin{equation}
\beta^{[2|1]} (\lambda) = -\frac{11}{3} \lambda^2 \frac{[1-2.87219
\lambda - 4.00209 \lambda^2]}{[1 - 5.96310 \lambda]}
\end{equation}
\begin{equation}
\beta^{[1|2]} (\lambda) = -\frac{11}{3} \lambda^2 \frac{[1-5.40935
\lambda]}{[1 - 8.50026 \lambda + 11.8442 \lambda^2]}.
\end{equation}
In both of these approximations, the first positive zero of the
denominator [0.1677 in (A.2) and 0.1483 in (A.3)] precedes the first
positive zero of the numerator [0.2595 in (A.2) and 0.1849 in (A.3)],
excluding Figure 1 type infrared-stable fixed point dynamics.  Rather,
both approximations support Figure 2 type dynamics, in which the first
positive denominator zero ($\lambda_d$) serves as an infrared attractor
for both a weak and a strong ultraviolet phase.

Such behaviour suggests that both phases may share common infrared
properties, as the running couplants in both phases evolve towards the
same infrared attractor.  This behaviour is corroborated by $[1|3]$ and
$[2|2]$ Pad\'e-summation $\beta$-functions whose Maclaurin expansions
reproduce (A.1a) inclusive of an arbitrary five-loop contribution
$R_4$.  For example, we find from comparison of (2.15) to (A.1a) that

\renewcommand{\theequation}{A.4\alph{equation}}
\setcounter{equation}{0}

\begin{equation}
\beta^{[1|3]} (\lambda) = - \frac{11}{3} \lambda^2 \left[ \frac{1+a_1
\lambda}{1+b_1 \lambda + b_2 \lambda^2 + b_3 \lambda^3} \right];
\end{equation}
\begin{equation}
a_1 = 15.8424 - 0.0379166 R_4,
\end{equation}
\begin{equation}
b_1 = 12.7515 - 0.0379166 R_4,
\end{equation}
\begin{equation}
b_2 = -53.8429 + 0.117197 R_4,
\end{equation}
\begin{equation}
b_3 = -103.614 + 0.184865 R_4.
\end{equation}
In Figure 19 we have plotted the first positive numerator and
denominator zeros of (A.4a) as a function of $R_4$, the unknown 
five-loop term.  A positive denominator zero $(\lambda_d)$
occurs over the entire range of $R_4$ and precedes the
numerator zero $(\lambda_n)$ when positive,
consistent with Figure 2-type dynamics.  For the $[2|2]$ case, we
find from (2.16) and (A.1) that

\renewcommand{\theequation}{A.5\alph{equation}}
\setcounter{equation}{0}

\begin{equation}
\beta^{[2|2]} (\lambda) = -\frac{11}{3} \lambda^2 \left[ \frac{1+a_1
\lambda + a_2 \lambda^2}{1+ b_1 \lambda + b_2 \lambda^2} \right];
\end{equation}
\begin{equation}
a_1 = 24.5902 - 0.0535237 R_4
\end{equation}
\begin{equation}
a_2 = -47.3211 + 0.0844278 R_4,
\end{equation}
\begin{equation}
b_1 = 21.4993 - 0.0535237 R_4,
\end{equation}
\begin{equation}
b_2 = -128.203 + 0.249865 R_4.
\end{equation}
As is evident from Fig. 20, the first positive numerator zero $(\lambda_n)$ is always 
preceded by a positive denominator zero $(\lambda_d)$, which serves as
the infrared-attractor for Figure 2-type dynamics, regardless of the
unknown five-loop term $R_4$.

Incorporation of a $[3|1]$ approximant within the Pad\'e-summation of
(A.1) yields a positive denominator zero only if $R_4 > 0$.  This is
because  $\lambda_d = R_3 / R_4$, as is evident from (2.16a) and
(2.16e).  Hence, if $R_4 < 0$, infrared dynamics along the lines of Fig.
2 are no longer possible.  However, we have found that no positive numerator zero exists
for this regime, also excluding the possibility of Fig. 1 type dynamics governed by 
an infrared-stable fixed point.  If $R_4$ is positive, the denominator
zero is once again seen to precede all positive numerator zeros, again
consistent with Figure 2 type dynamics.  All of this behaviour can be extracted
from (2.16) and (A.1):

\renewcommand{\theequation}{A.6\alph{equation}}
\setcounter{equation}{0}

\begin{equation}
\beta^{[3|1]} (\lambda) = -\frac{11}{3} \lambda^2 \frac{[1+a_1 \lambda +
a_2 \lambda^2 + a_3 \lambda^3]}{[1+b_1 \lambda]}
\end{equation}
\begin{equation}
a_1 = 3.09091-0.0116221 R_4
\end{equation}
\begin{equation}
a_2 = 14.2929-0.0359227 R_4,
\end{equation}
\begin{equation}
a_3 = 86.0433-0.167698 R_4,
\end{equation}
\begin{equation}
b_1 = -0.0116221 R_4.
\end{equation}
The absence of a positive numerator zero when $R_4$ is negative
necessarily follows from the fact that $a_1, a_2$, and $a_3$ are all
positive when $R_4 < 0$.

\newpage

\begin{figure}
\includegraphics[scale=0.7,angle=270]{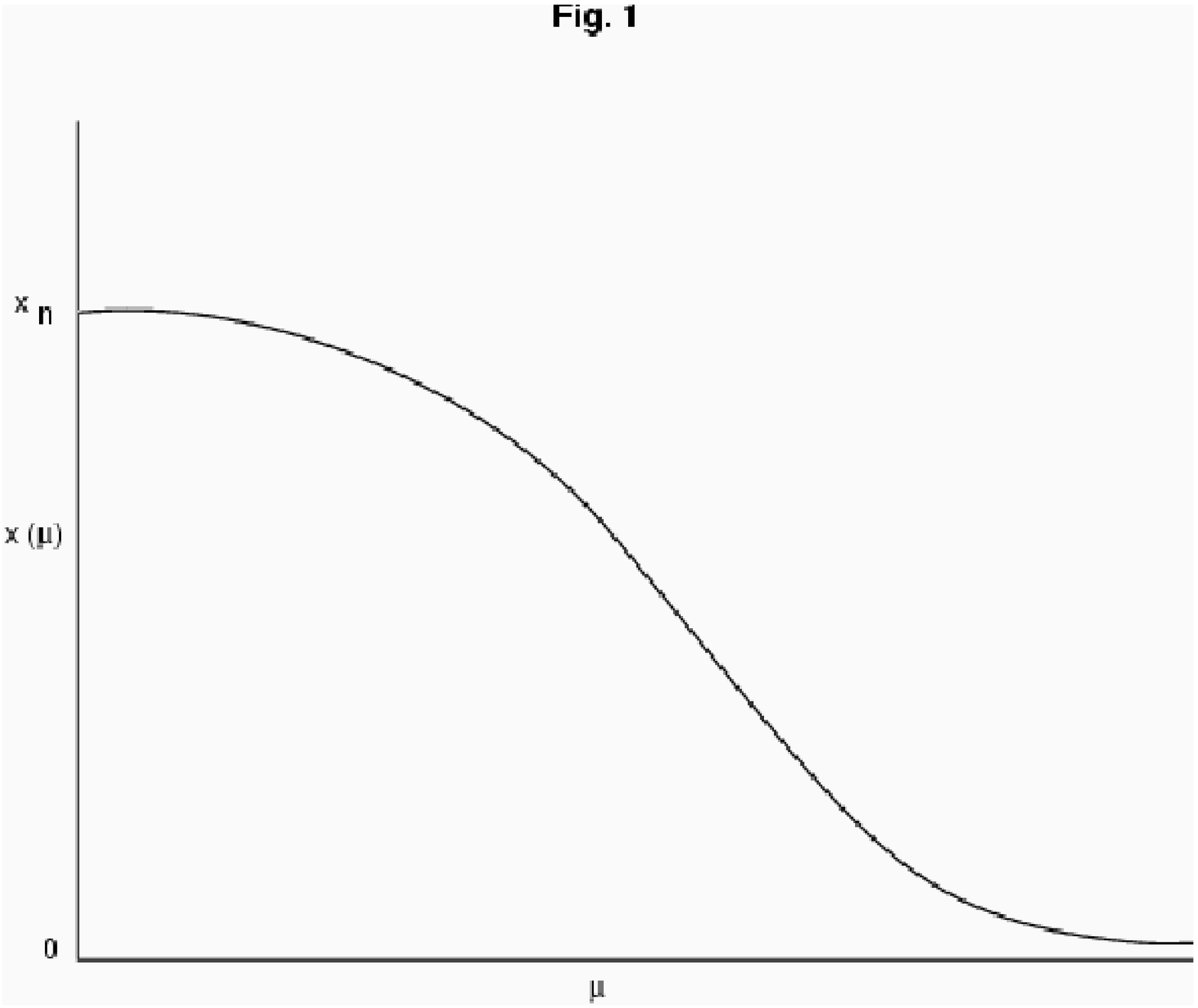}
\caption{
Behaviour of the asymptotically-free running couplant
$x(\mu)\equiv\alpha_s (\mu)/\pi$ in dynamics governed by an infrared-stable 
fixed point $x_n$.
}
\end{figure}

\clearpage

\begin{figure}
\includegraphics[scale=0.7,angle=270]{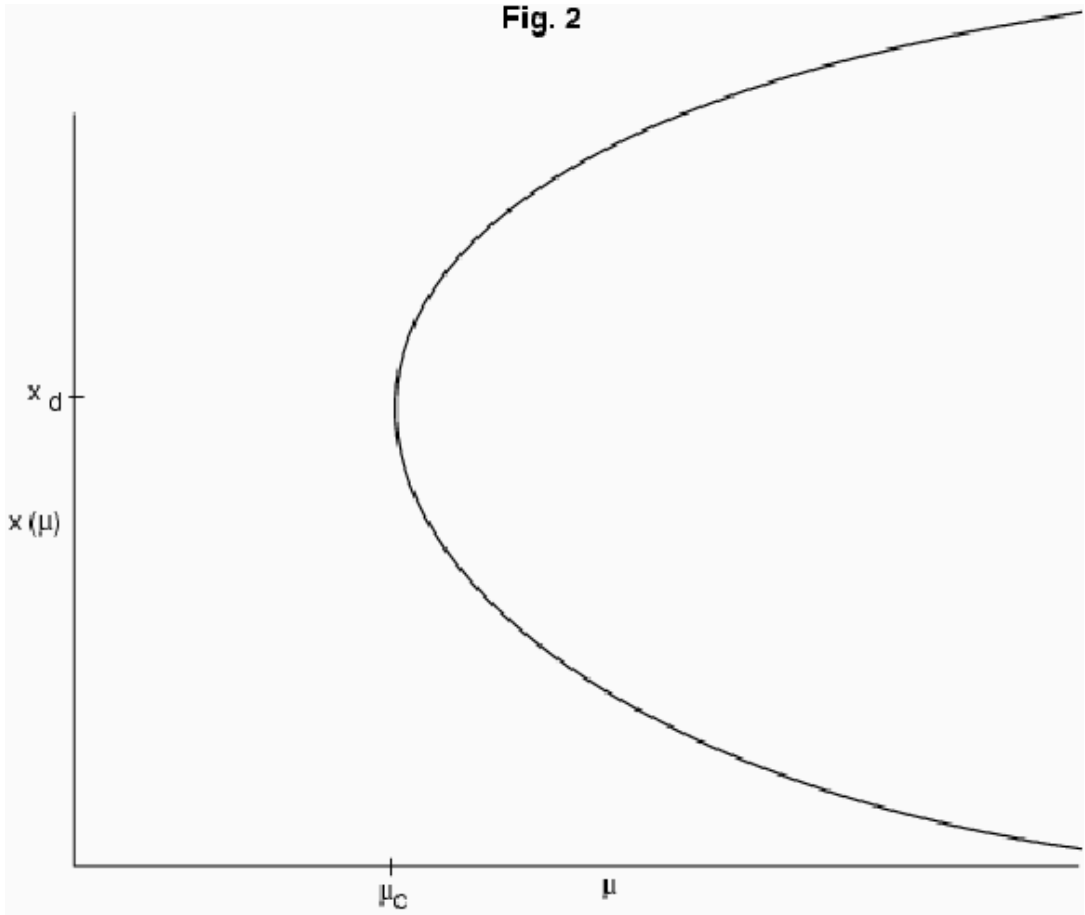}
\caption{
 Behaviour of the asymptotically-free running couplant $x(\mu)$ in dynamics governed by a
$\beta$-function pole at $x_d$.  The point $x_d$ serves as an infrared attractor of both a strong and weak phase
of the couplant. An infrared cut-off $\mu_c$ necessarily occurs, corresponding to the $\beta$-function pole at
$x(\mu_c) = x_d$.
}
\end{figure}

\clearpage

\begin{figure}
\includegraphics[scale=0.7,angle=270]{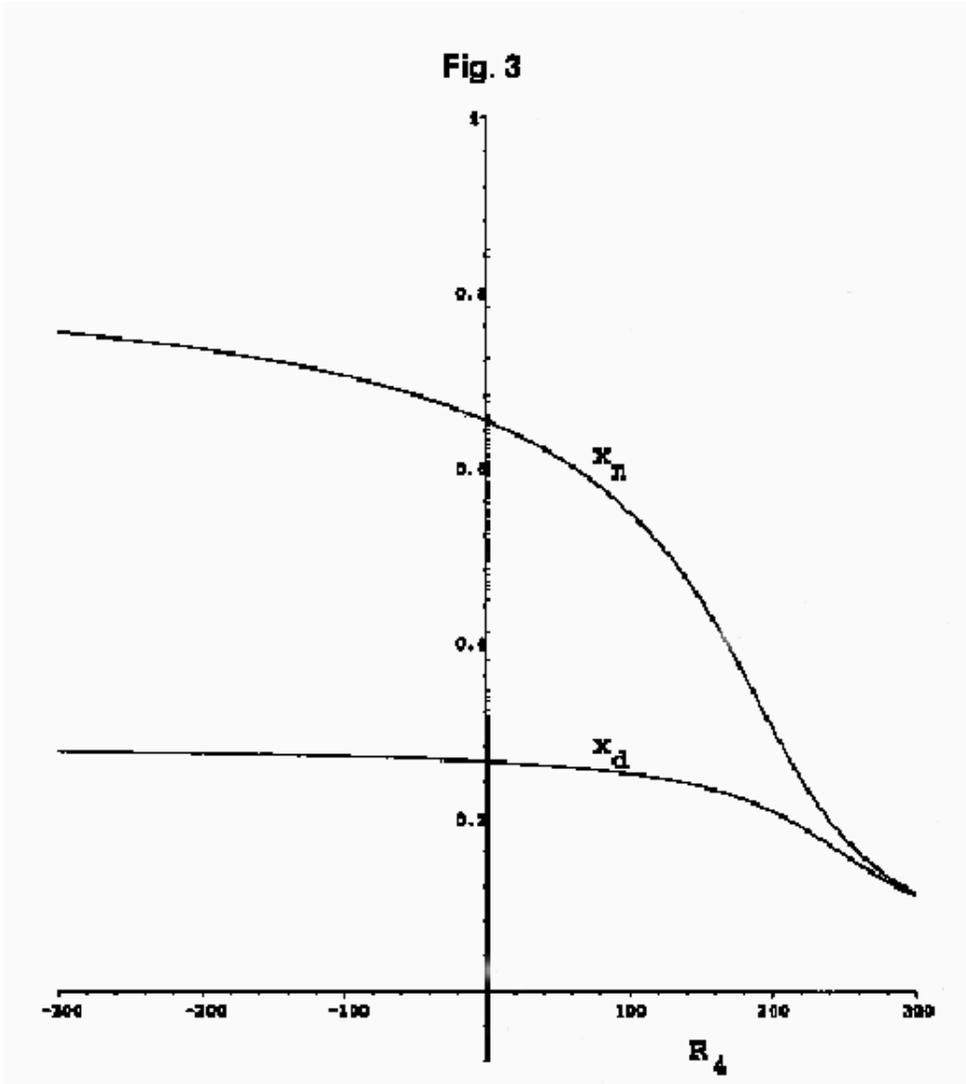}
\caption{
 The $R_4$-dependence of the first positive numerator zero $(x_n)$ and the first positive
denominator zero $(x_d)$ of the $[2|2]$ Pad\'e-summation of the QCD $\beta$-function when $n_f = 0$. The
independent variable $R_4 \equiv \beta_4/\beta_0$ is proportional to the presently unknown 5-loop contribution $(\beta_4)$ 
to the $\beta$-function. The figure shows that $x_d$ and $x_n$ both
exist over the entire range of $R_4$. However, $x_d$ always precedes $x_n$, indicative of the dynamics of Figure 2 
for the evolution of the couplant $x(\mu)$ from the asymptotically-free ultraviolet region.
}
\end{figure}

\clearpage

\begin{figure}
\includegraphics[scale=0.7,angle=270]{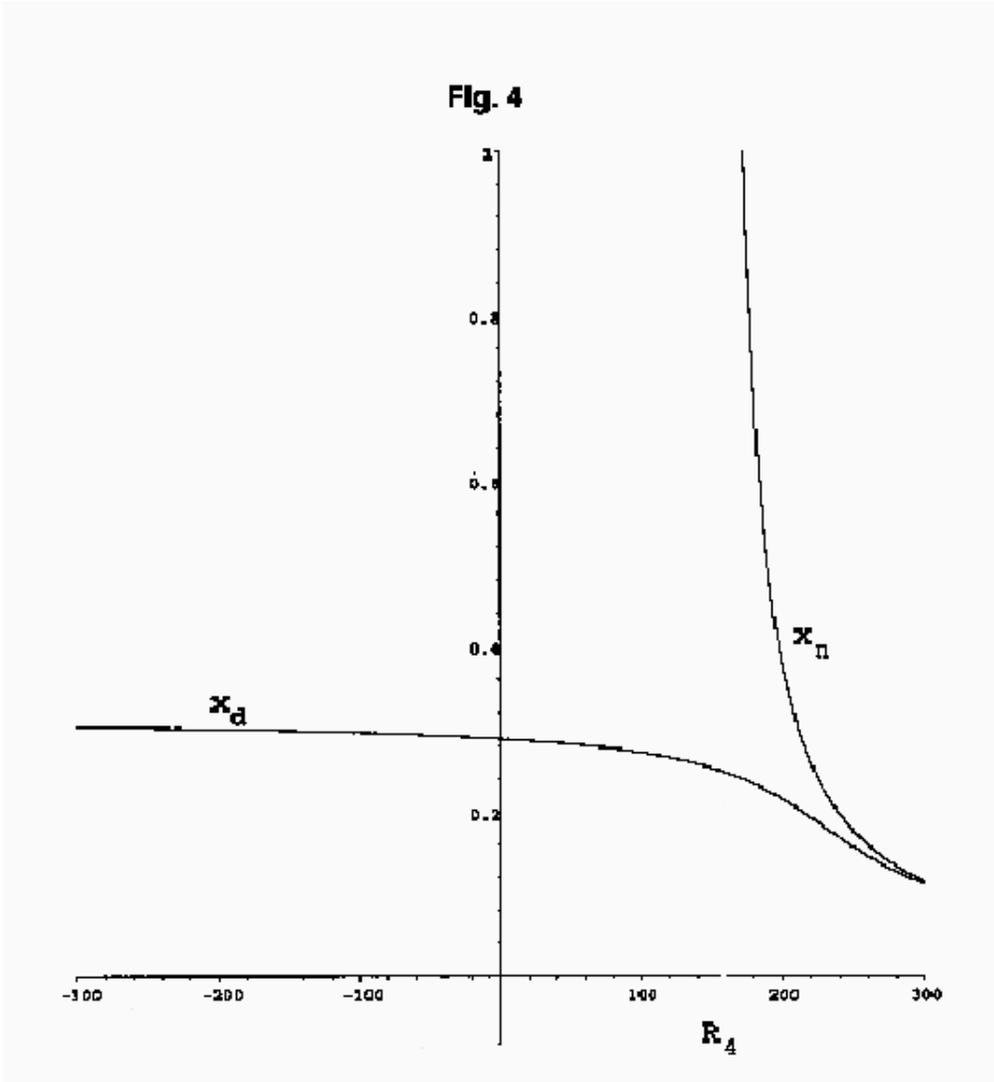}
\caption{
The $R_4$-dependence $[R_4 \equiv \beta_4/\beta_0]$ of the first positive numerator zero $(x_n)$ and the 
first positive denominator zero $(x_d)$ of the $[1|3]$ Pad\'e-summation of the QCD $\beta$-function when $n_f = 0$.  The
figure shows that $x_d$ exists over the entire range of $R_4$. Moreover, $x_d$ precedes $x_n$ over the range of
$R_4$ for which a positive numerator zero exists, leading once again to the dynamics of Figure 2 for the
evolution of the couplant from the asymptotically-free ultraviolet region.
}
\end{figure}

\clearpage

\begin{figure}
\includegraphics[scale=0.7,angle=270]{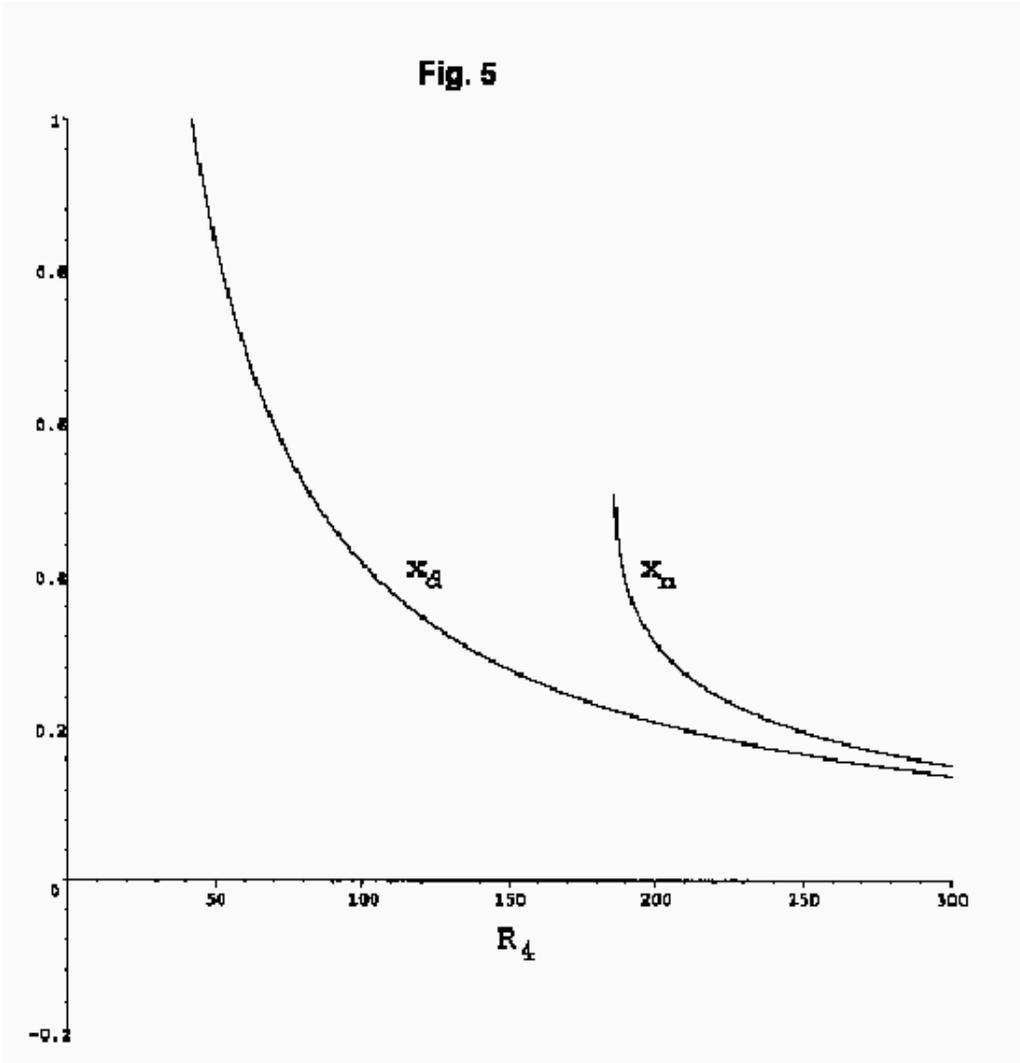}
\caption{
The $R_4$-dependence $[R_4 \equiv \beta_4/\beta_0]$ of the first positive numerator zero $(x_n)$ and the first
positive denominator zero $(x_d)$ of the $[3|1]$ Pad\'e-summation of the QCD $\beta$-function when $n_f = 0$.  As in
Figures 3 and 4, $x_d$ exists over the entire range of $R_4$, and $x_d$ precedes $x_n$ where a positive numerator
zero exists.
}
\end{figure}

\clearpage

\begin{figure}
\includegraphics[scale=0.7,angle=270]{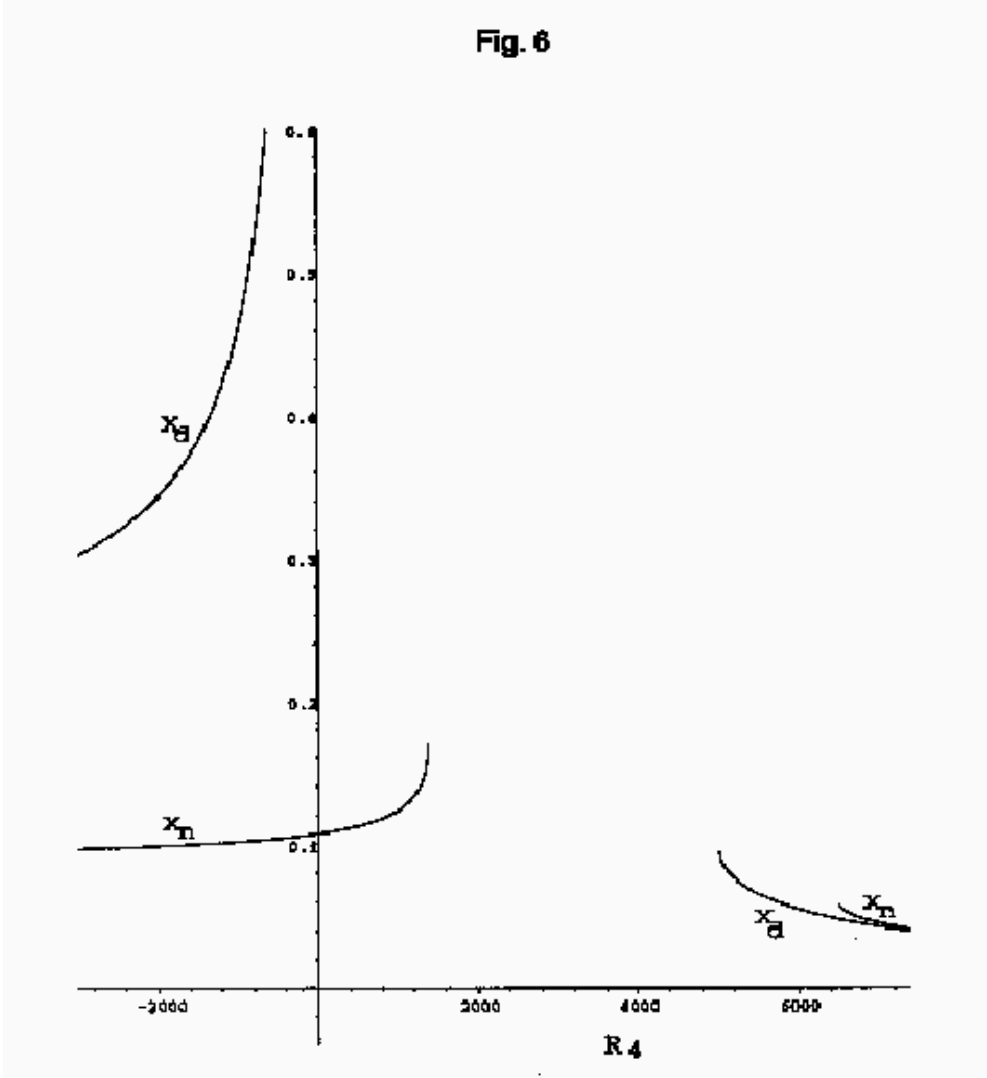}
\caption{
The behaviour over a very large region of $R_4$ of the first positive numerator zero $(x_n$) and
the first positive denominator zero $(x_d)$ of the $[2|2]$ Pad\'e-summation of the QCD $\beta$-function when
$n_f = 13$. The figure shows that $x_n$ precedes $x_d$ when $R_4 < 1383$, indicative of dynamics governed by
an infrared-stable fixed point (Figure 1) for this region.  The figure also shows that $x_d$ precedes $x_n$
when $R_4$ is very large ($R_4 > 5020$), indicative of dynamics governed by a $\beta$-function pole (Figure 2).
The absence of a positive $\beta$-function zero for $R_4$ between 1383 and 5020 precludes 
the possibility of dynamics governed by an infrared-stable fixed point for this 
range of $R_4$. 
}
\end{figure}

\clearpage

\begin{figure}
\includegraphics[scale=0.7,angle=270]{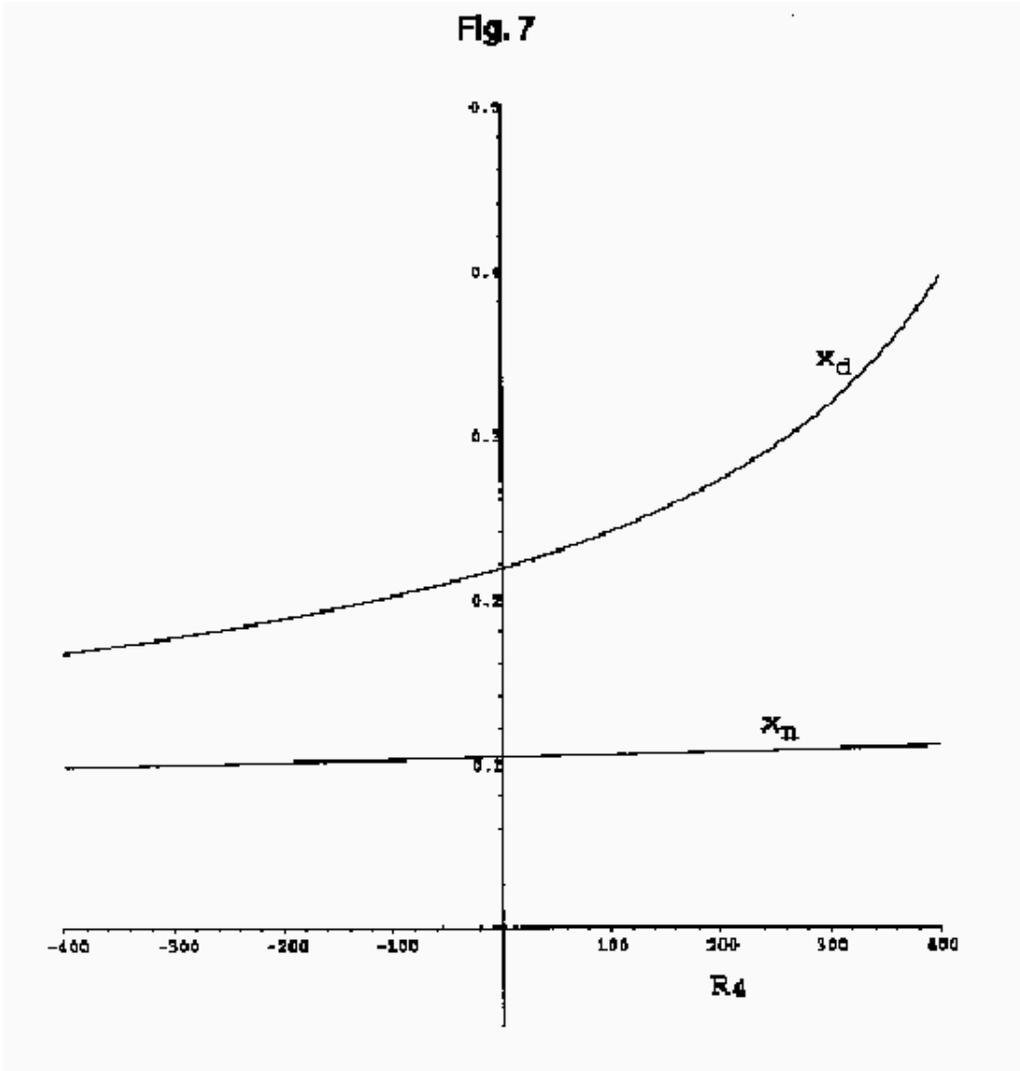}
\caption{
The $R_4$-dependence of the first positive numerator zero $(x_n)$ and the first positive
denominator zero $(x_d)$ of the $[1|3]$ Pad\'e-summation of the QCD $\beta$-function when $n_f = 13$. The
numerator zero is seen to precede the denominator zero over the entire range of $R_4$ exhibited,
consistent with the numerator zero serving as an infrared-stable fixed point. This numerator zero
ceases to be positive when $R_4 > 6394$.
}
\end{figure}

\clearpage

\begin{figure}
\includegraphics[scale=0.7,angle=270]{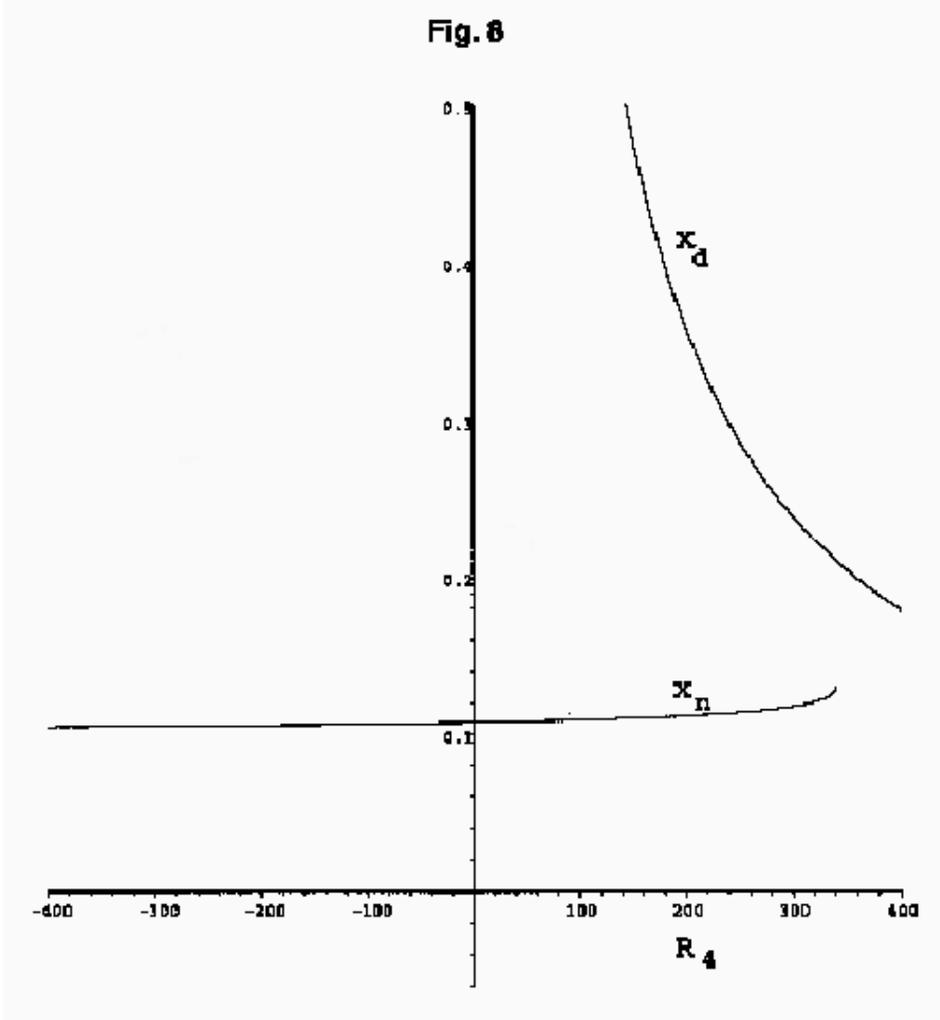}
\caption{
The $R_4$-dependence of the first positive numerator zero $(x_n)$ and the first positive
denominator zero $(x_d)$ of the $[3|1]$ Pad\'e-summation of the QCD $\beta$-function when $n_f = 13$. The
numerator zero is seen to precede the denominator zero when $R_4 < 340$, corresponding to the domain
of $R_4$ for dynamics governed by the infrared-stable fixed point $x_n$. The denominator zero precedes
any positive numerator zeros that occur for $R_4 > 340$, corresponding to Figure 2 dynamics governed
by an infrared attractor at $x_d$.
}
\end{figure}

\clearpage

\begin{figure}
\includegraphics[scale=0.7,angle=270]{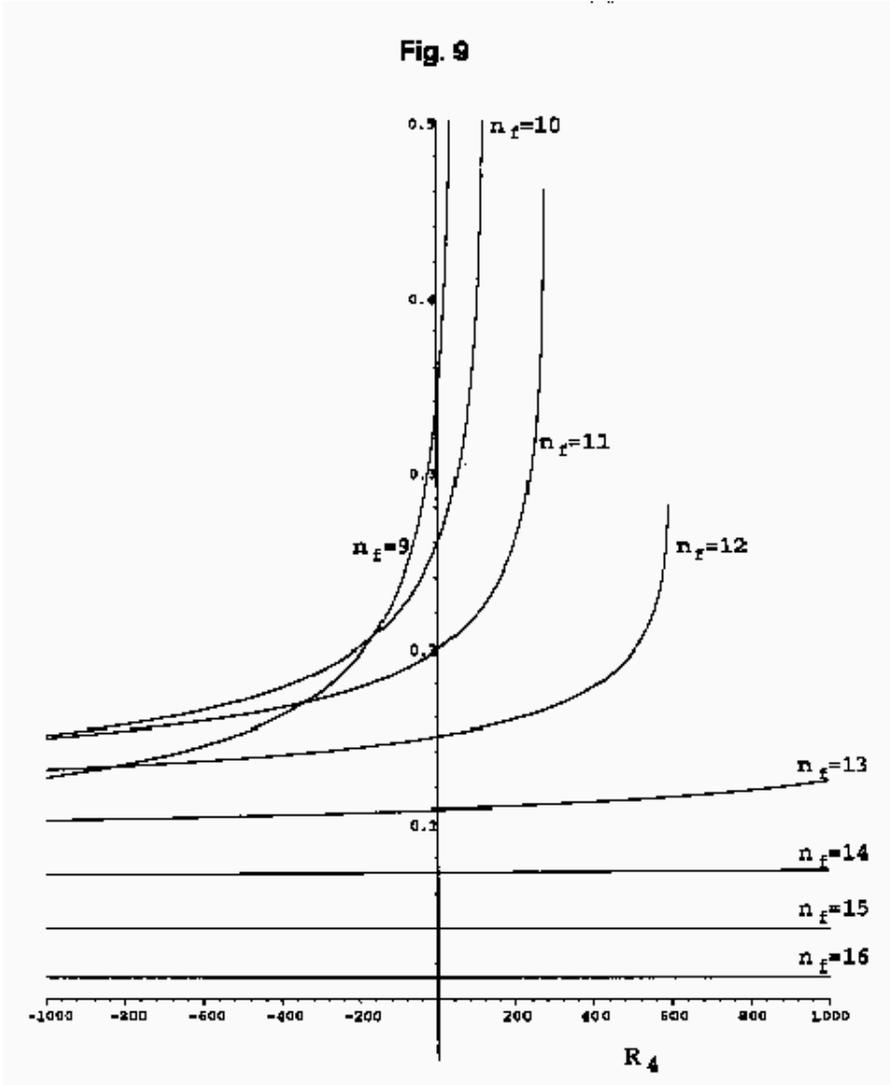}
\caption{
 The $n_f$-dependence of the infrared fixed point $x_n$, as indicated by successive plots of $x_n$
versus $R_4$ for $\beta^{[2|2]}$ with values of $n_f$ ranging from 9-16.
}
\end{figure}

\clearpage

\begin{figure}
\includegraphics[scale=0.7,angle=270]{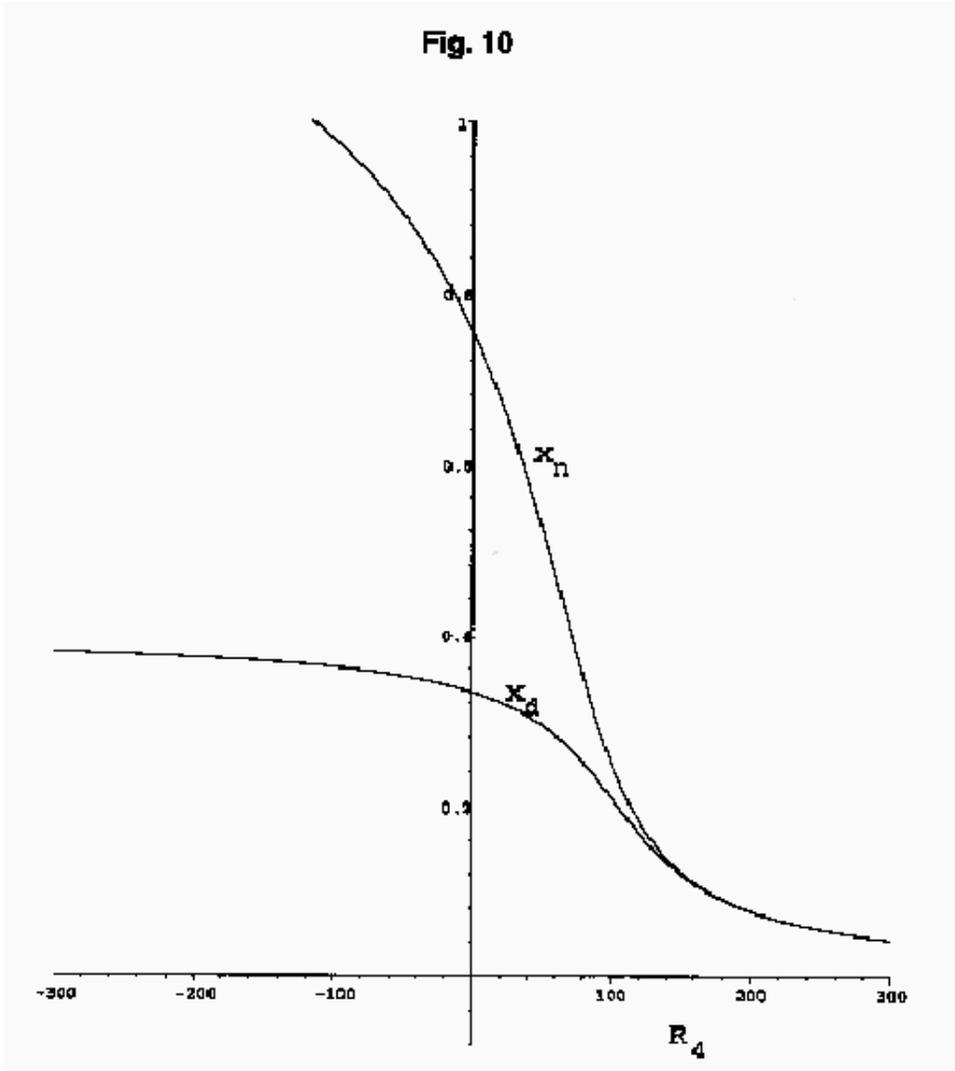}
\caption{
The $R_4$-dependence of the first positive numerator zero $(x_n)$ and the first positive
denominator zero $(x_d)$ of the $[2|2]$ Pad\'e-summation of the QCD $\beta$-function when $n_f = 3$. The figure
shows that $x_d$ always precedes $x_n$, indicative of infrared-attractor dynamics (Figure 2).
}
\end{figure}

\clearpage

\begin{figure}
\includegraphics[scale=0.7,angle=270]{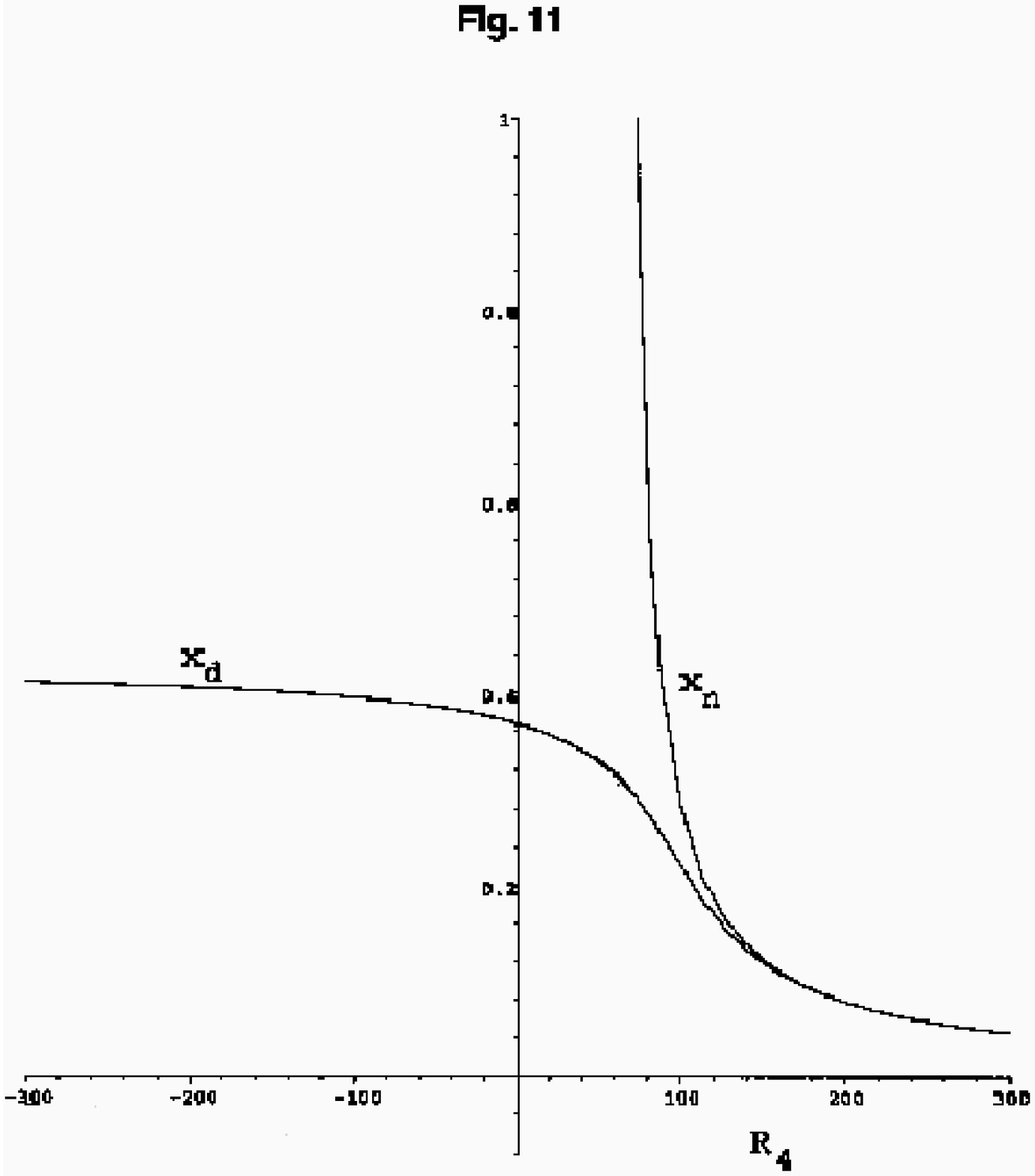}
\caption{
The $R_4$-dependence of the first positive numerator zero $(x_n)$ and the first positive
denominator zero $(x_d)$ of the $[1|3]$ Pad\'e-summation of the QCD $\beta$-function when $n_f = 3$. The figure
shows that $x_d$ exists over the entire range of $R_4$, and precedes $x_n$ over the range where such a positive
zero exists. Hence $x_d$ serves as an infrared attractor.
}
\end{figure}

\clearpage

\begin{figure}
\includegraphics[scale=0.7,angle=270]{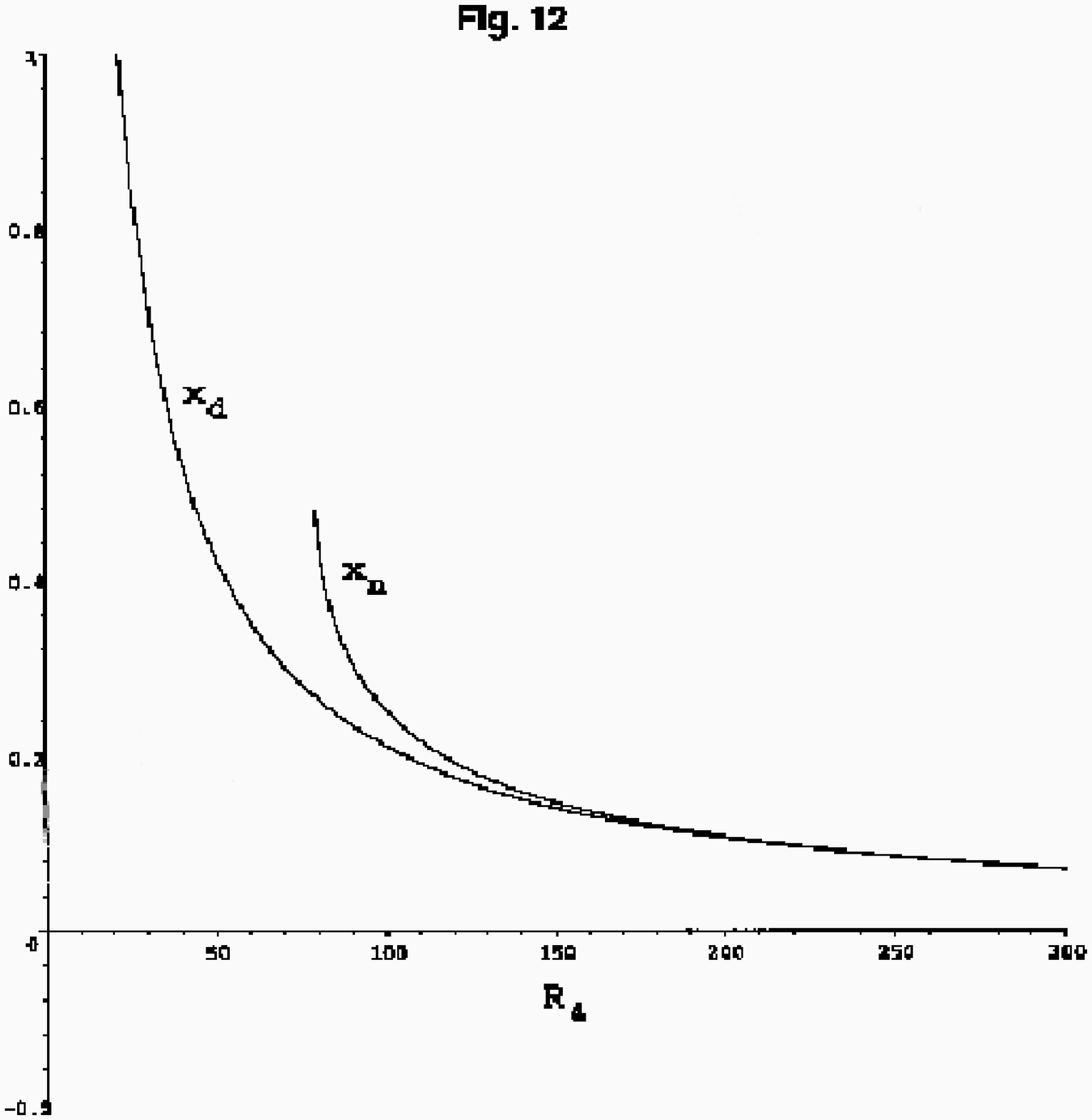}
\caption{
 The $R_4$-dependence of the first positive numerator zero $(x_n)$ and the first positive
denominator zero $(x_d)$ of the $[3|1]$ Pad\'e-summation of the QCD $\beta$-function when $n_f = 3$. The figure
shows that $x_d$ exists over the entire range of $R_4$, and precedes $x_n$ over the range where such a positive
zero exists, as in Fig. 11. Hence $x_d$ serves as an infrared attractor for this case as well.
}
\end{figure}

\clearpage

\begin{figure}
\includegraphics[scale=0.7,angle=270]{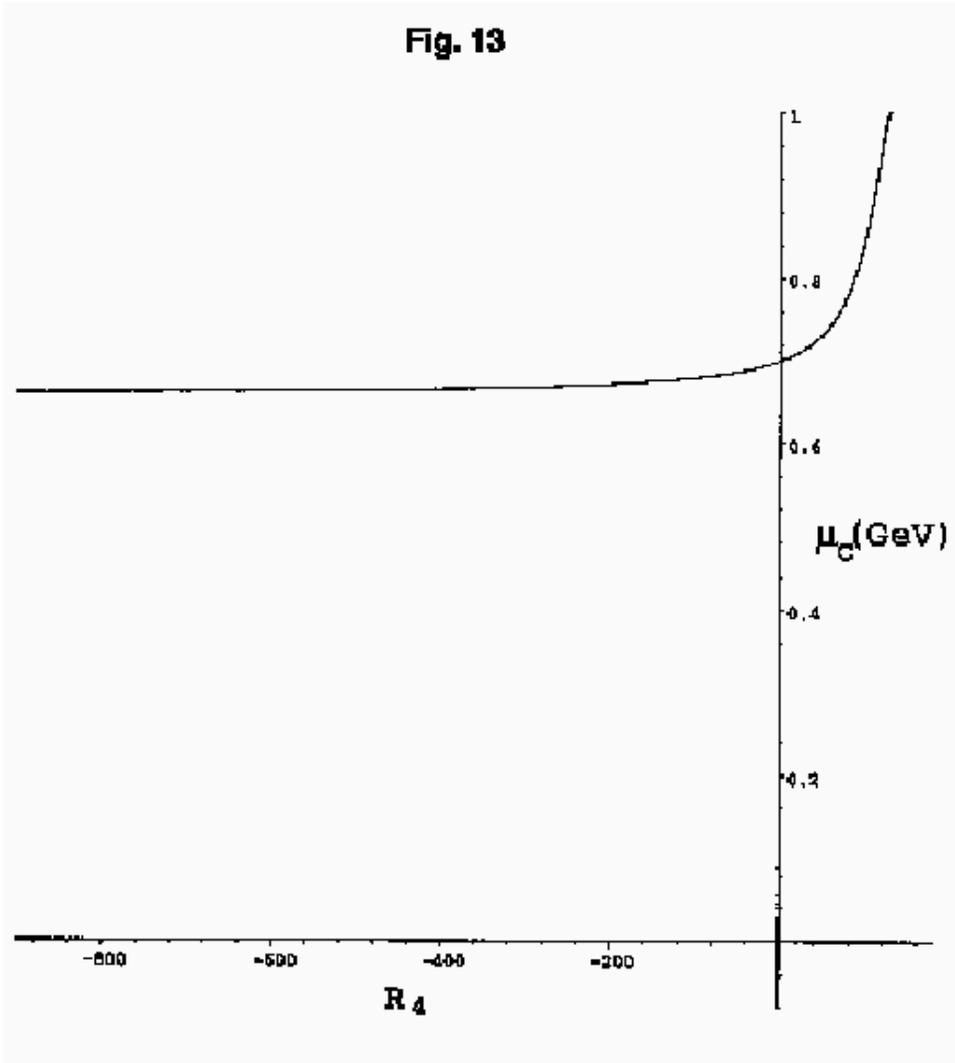}
\caption{
The $R_4$-dependence of $\mu_c$, the minimum value of $\mu$ accessible to the couplant $x(\mu)$
derived from $\beta^{[2|2]}$ when $n_f = 3$.  The 1 GeV value of the couplant is assumed to be x(1) = 0.153, as
discussed in the text
}
\end{figure}

\clearpage

\begin{figure}
\includegraphics[scale=0.7,angle=270]{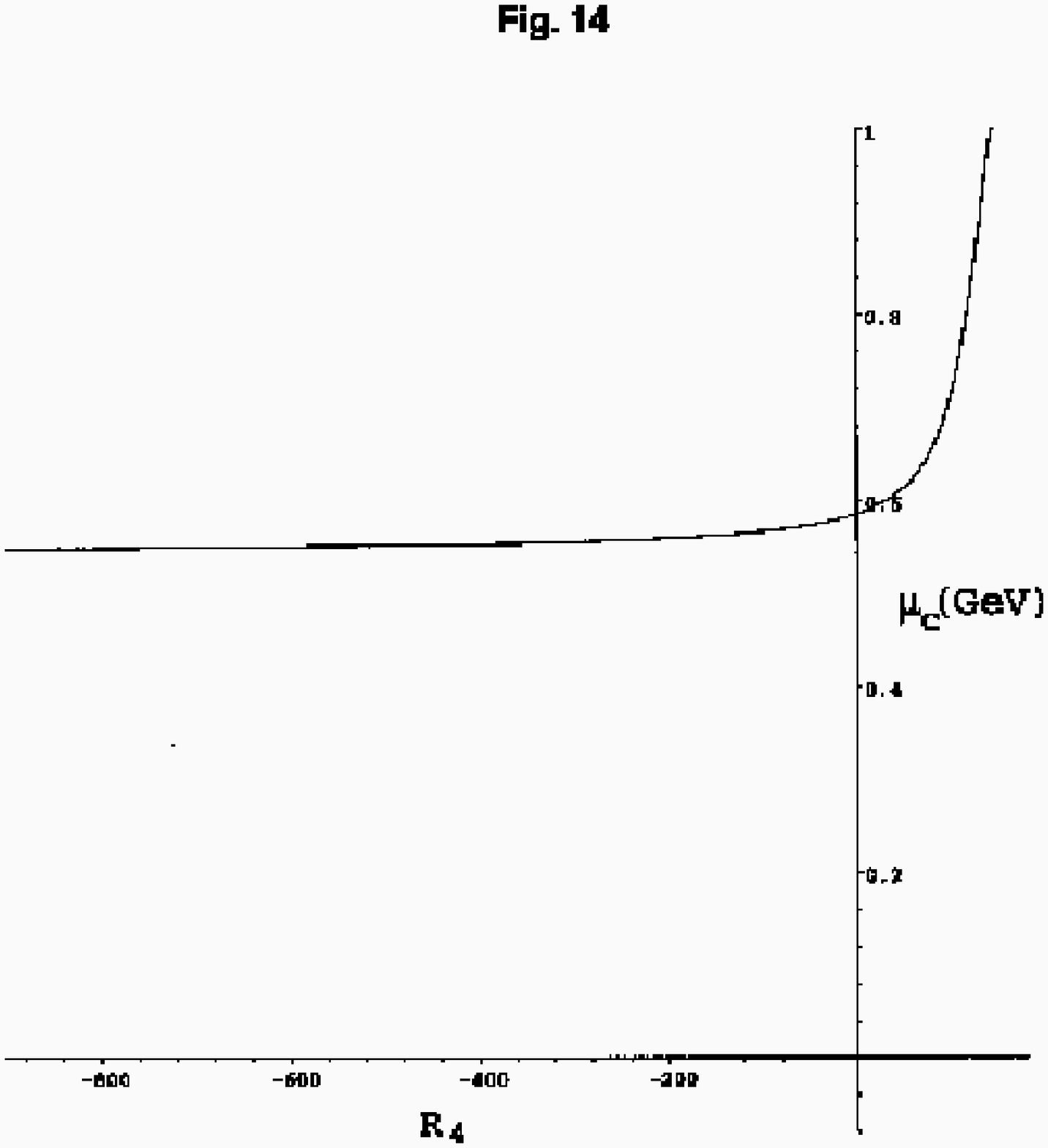}
\caption{
The $R_4$-dependence of $\mu_c$, as in Figure 13, except that the 1 GeV value of the couplant
is given a lower-bound value $x(1)$ = 0.1305.
}
\end{figure}

\clearpage

\begin{figure}
\includegraphics[scale=0.7,angle=270]{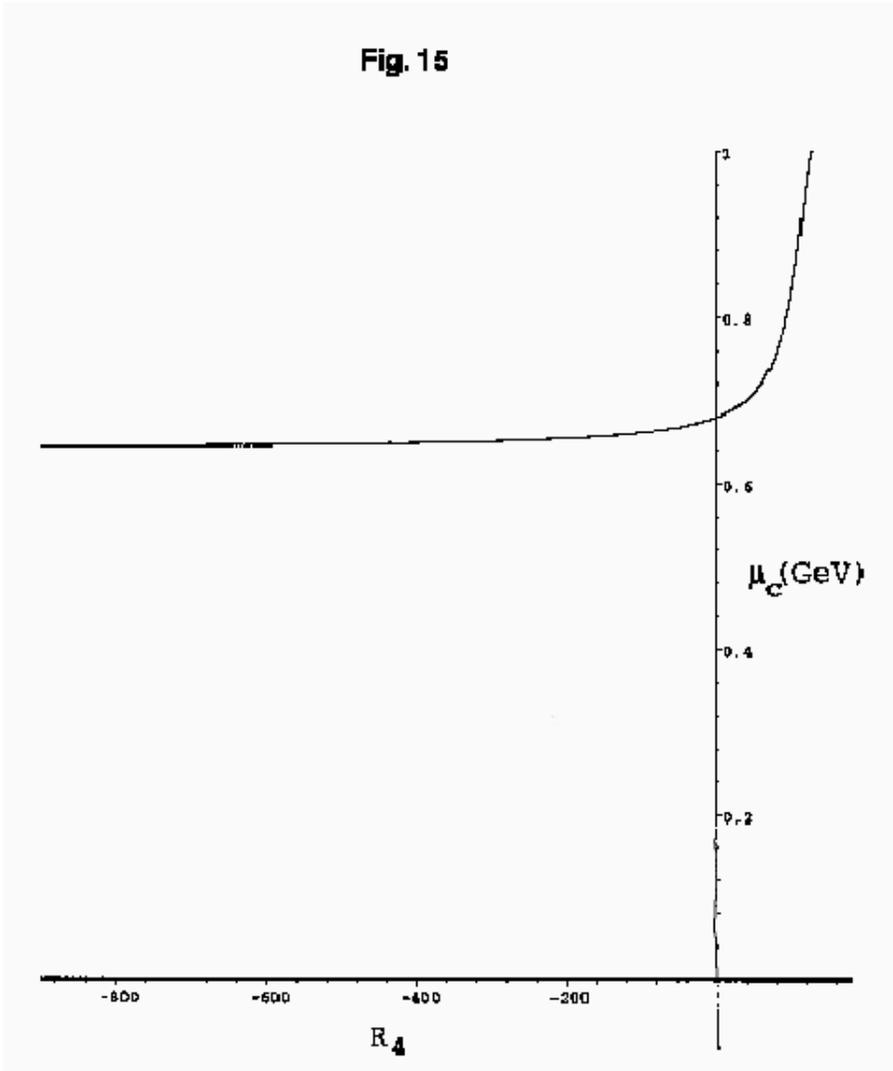}
\caption{
 The $R_4$-dependence of $\mu_c$, the minimum value of $\mu$ accessible to the couplant $x(\mu)$
derived from $\beta^{[1|3]}$ when $n_f = 3$ and $x(1)$ = 0.153.
}
\end{figure}

\clearpage

\begin{figure}
\includegraphics[scale=0.7,angle=270]{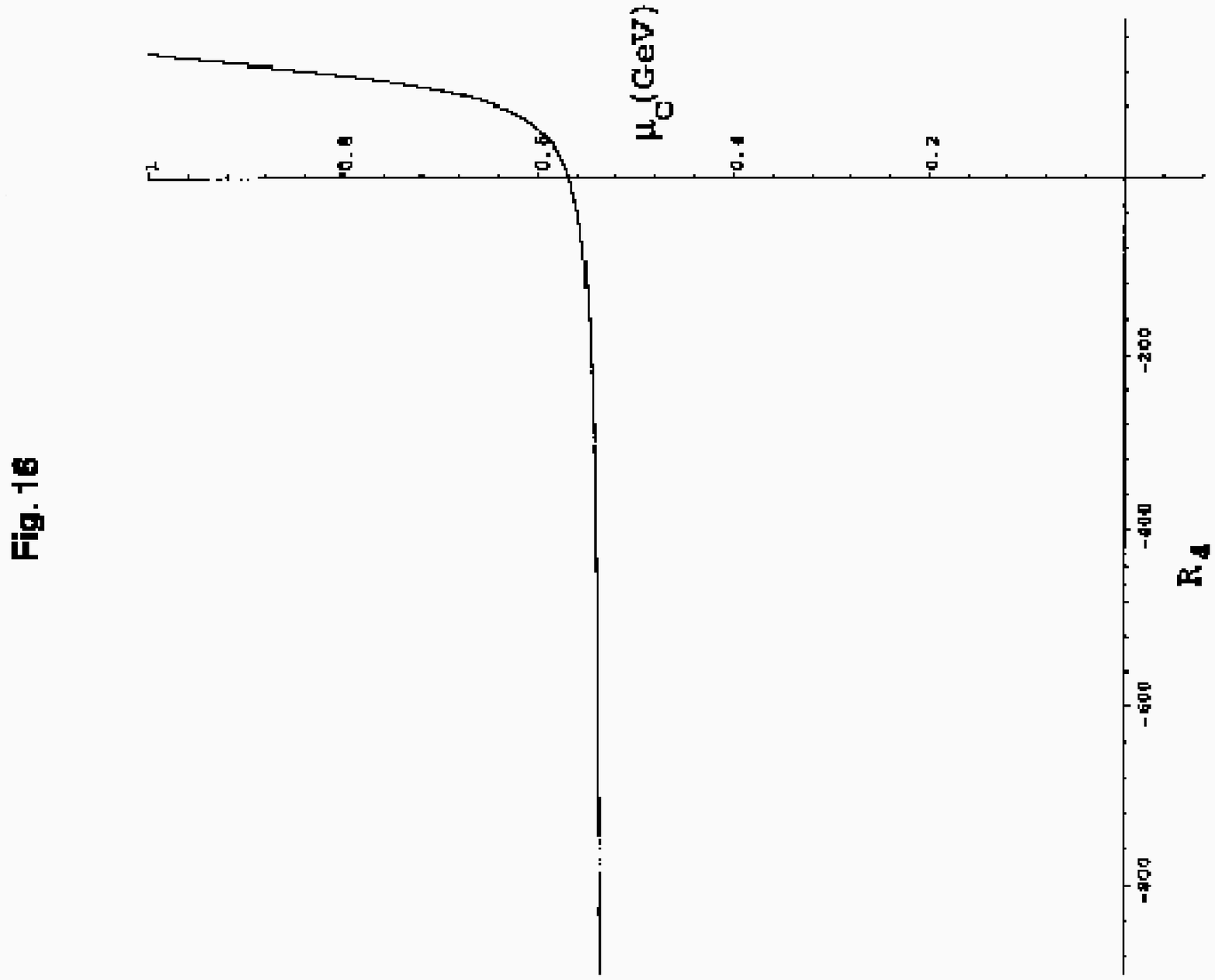}
\caption{
 The $R_4$-dependence of $\mu_c$, as in Figure 15, except that the 1 GeV value of the couplant
is given a lower-bound value $x(1)$ = 0.1305.
}
\end{figure}

\clearpage

\begin{figure}
\includegraphics[scale=0.7,angle=270]{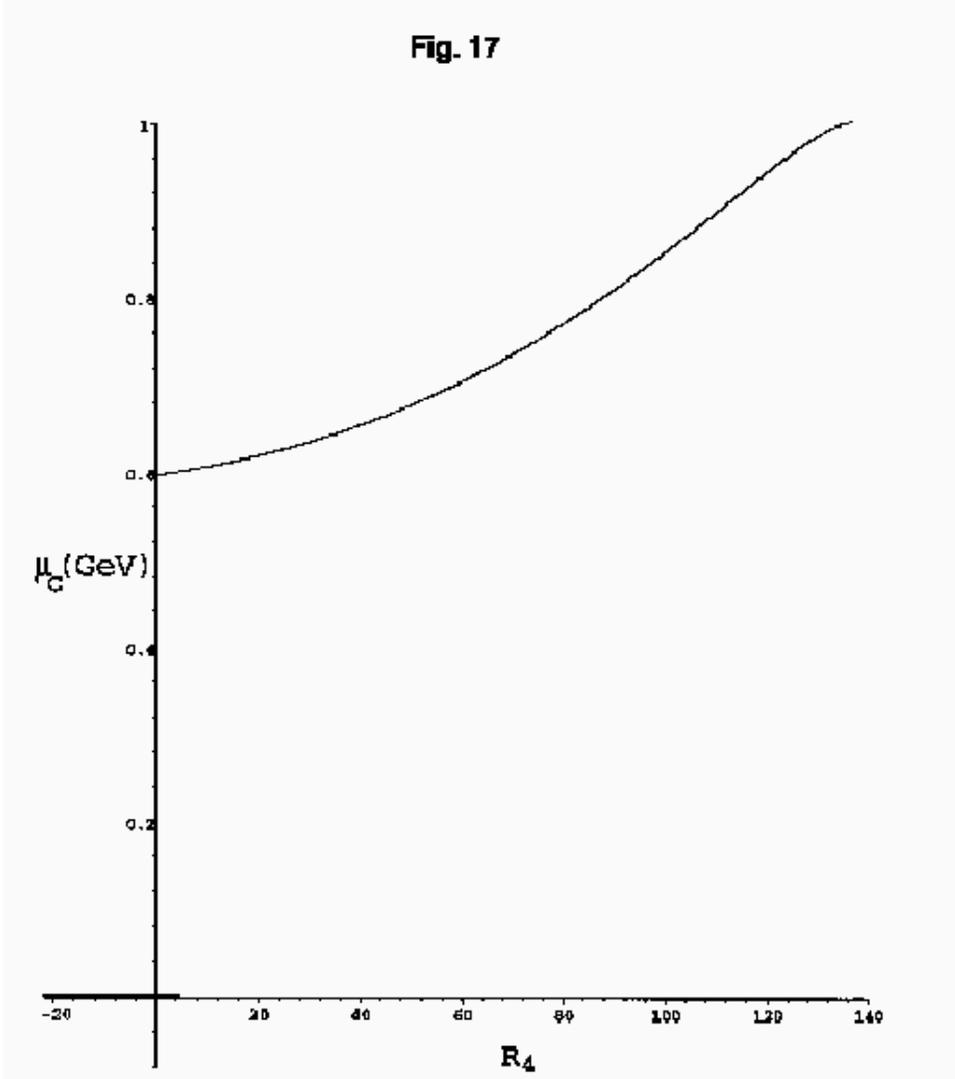}
\caption{
 The $R_4$-dependence of $\mu_c$, the minimum value of $\mu$ accessible to the couplant $x(\mu)$
derived from $\beta^{[3|1]}$ when $n_f = 3$ and $x(1)$ = 0.153. The couplant has an infrared attractor for this case
only if $R_4$ is positive.
}
\end{figure}

\clearpage

\begin{figure}
\includegraphics[scale=0.7,angle=270]{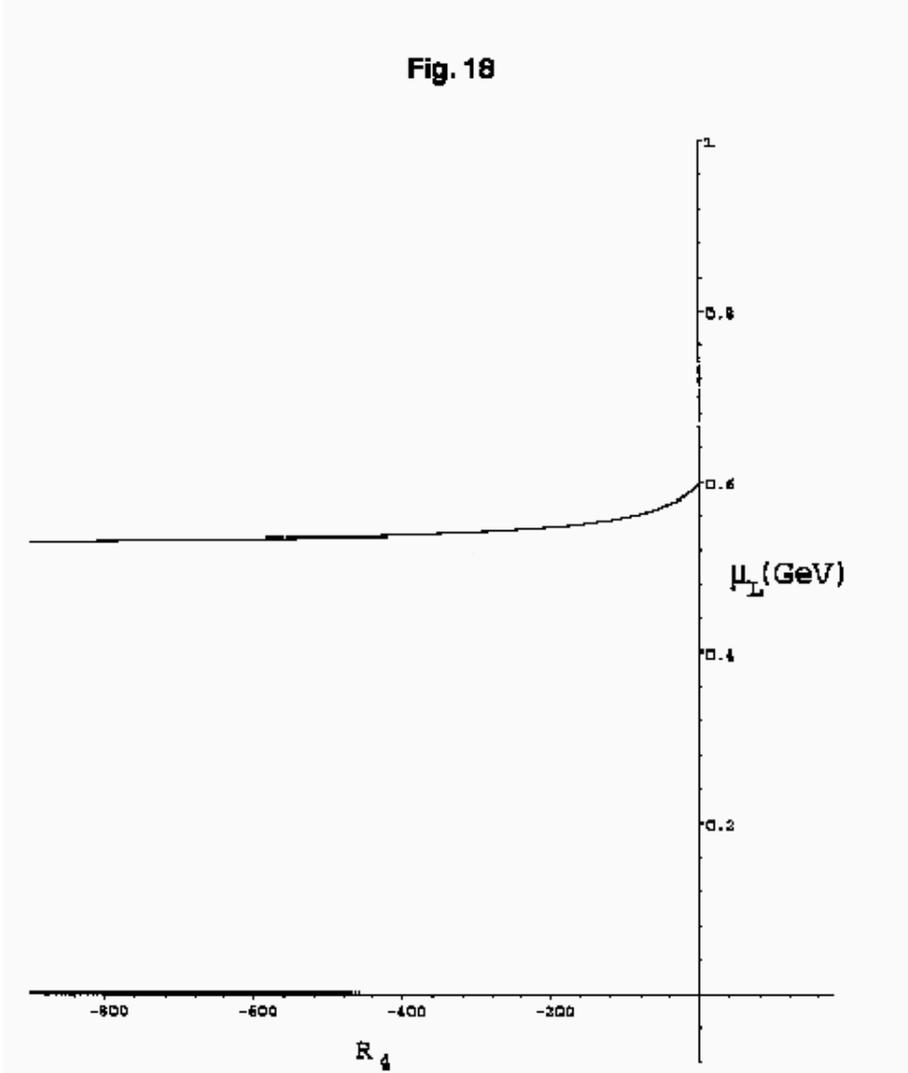}
\caption{
 The $R_4$-dependence of the Landau pole $\mu_L$ of the couplant derived from $\beta^{[3|1]}$ when $n_f =
3$ and $x(1)$ = 0.153. This Landau pole (as opposed to an infrared attractor of two ultraviolet phases)
occurs only for negative values of $R_4$.
}
\end{figure}

\clearpage

\begin{figure}
\includegraphics[scale=0.7,angle=270]{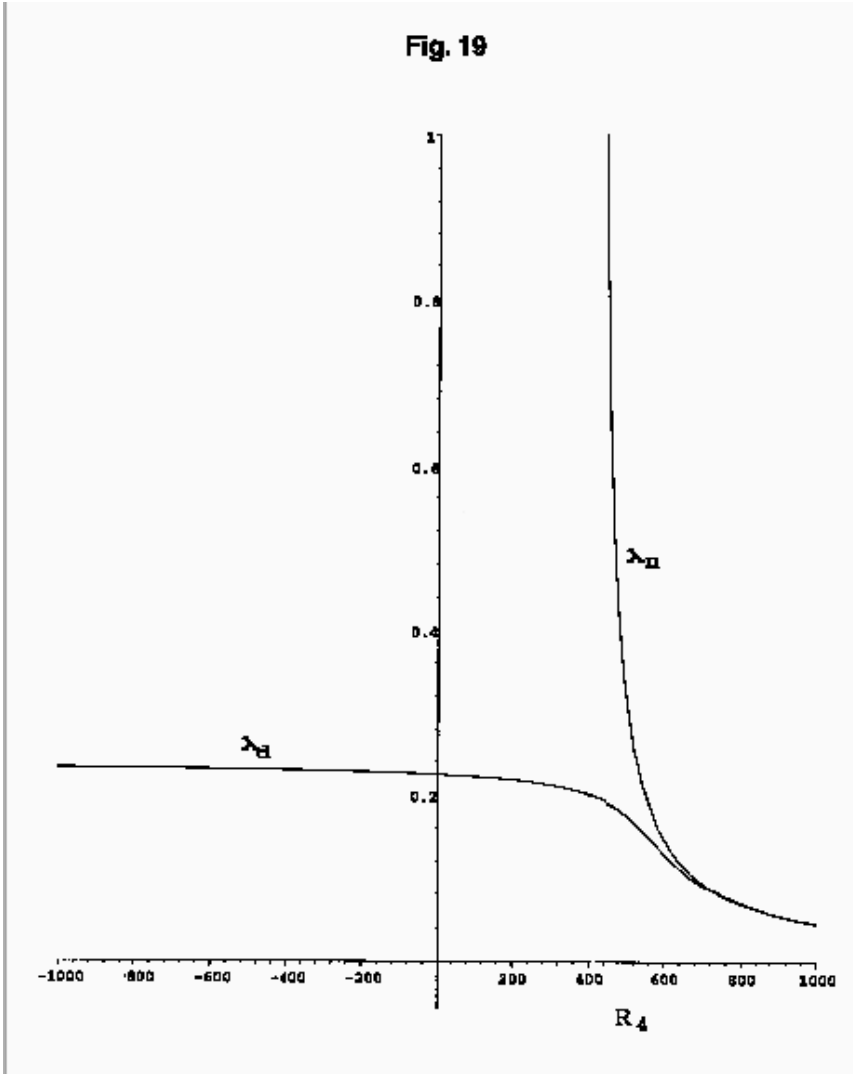}
\caption{
The dependence on the unknown five-loop term $R_4$ of the first positive
numerator zero ($\lambda_n$) and the first positive denominator zero ($\lambda_d$) of the $[1|3]$
Pad\'e-summation of the $SU(N_c)$ $\beta$-function for the couplant $\lambda = N_c \alpha(\mu) / 4 \pi$ in the
$N_c \rightarrow \infty$ limit ($n_f = 0$).  The figure shows that the denominator zero $\lambda_d$
exists over the range of $R_4$ and precedes the first positive numerator zero $\lambda_n$ when such a zero
exists, indicative of Figure 2 dynamics with $\lambda_d$ serving as the infrared attractor.
}
\end{figure}

\clearpage

\begin{figure}
\includegraphics[scale=0.7,angle=270]{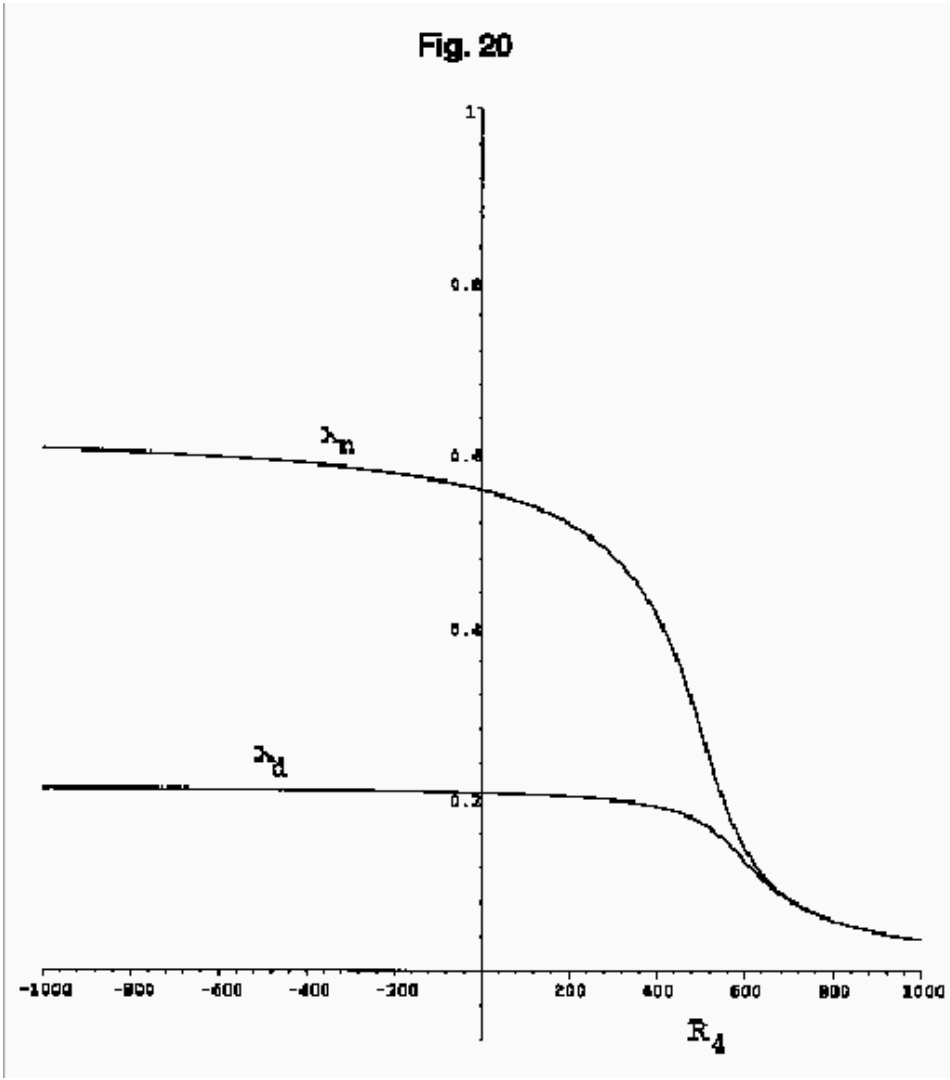}
\caption{
The $R_4$-dependence of the first positive numerator zero ($\lambda_n$) and the first
positive denominator zero ($\lambda_d$) of the $[2|2]$ Pad\'e-summation of the $SU(N_c)$ $\beta$-function for the couplant $\lambda
= N_c \alpha (\mu) / 4 \pi$ in the $N_c \rightarrow \infty$ limit ($n_f = 0$).  Once again, the denominator
zero $\lambda_d$ exists over the entire range of $R_4$ and precedes the first positive numerator zero $\lambda_n$.
}
\end{figure}

\end{document}